\documentclass[twocolumn,pra,aps,showpacs,superscriptaddress]{revtex4}

\usepackage{bbold}
\usepackage{mathptmx}
\usepackage{subfigure}
\usepackage{psfrag,graphicx}
\usepackage{dcolumn}
\usepackage{amsmath,amssymb}
\usepackage{bm}
\usepackage{color}
\usepackage{latexsym}
\usepackage{epstopdf}
\usepackage{color}
\usepackage[english]{babel}
\usepackage{latexsym}
\usepackage{psfrag,graphicx}
\usepackage{subfigure}
\usepackage{amsmath}
\usepackage{amssymb}
\usepackage{amsfonts}
\usepackage{bm}
\usepackage{natbib}
\usepackage{epstopdf}
\usepackage{appendix}

\definecolor{mygrey}{gray}{0.35}
\definecolor{myblue}{rgb}{0.2,0.2,0.8}
\definecolor{myzard}{cmyk}{0,0,0.05,0}
\definecolor{mywhite}{rgb}{1,1,1}
\definecolor{mywhite}{rgb}{1,1,1}
\definecolor{myred}{rgb}{1,0.,0.3}

\usepackage[colorlinks=true,citecolor=myblue,linkcolor=myred]{hyperref}

\def\ba{\begin{align}}
\def\enda{\end{align}}
\def\bi{\begin{itemize}}
\def\ei{\end{itemize}}

\def\be{\begin{equation}}
\def\ee{\end{equation}}
\def\bea{\begin{eqnarray}}
\def\eea{\end{eqnarray}}
\def\bse{\begin{subequations}}
\def\ese{\end{subequations}}

\newcommand{\ket}[1]{|{#1}\rangle}                       
\newcommand{\bra}[1]{\langle {#1}|}                      
\newcommand{\average}[1]{\langle {#1} \rangle}           

\newcommand{\Ignore}[1]{ }

\def\black{}

\def\magenta{\color{magenta}}

\begin{document}
\title{Spin-1/2 sub-dynamics nested in the quantum dynamics of two coupled qutrits}
\author{R. Grimaudo}
 \affiliation{ Dipartimento di Fisica e Chimica dell'Universit\`a di Palermo, Via Archirafi, 36, I-90123 Palermo, Italy}
%
\author{A. Messina}
\affiliation{ Dipartimento di Fisica e Chimica dell'Universit\`a di Palermo, Via Archirafi, 36, I-90123 Palermo, Italy}
\affiliation{ I.N.F.N., Sezione Catania}


\author{P. A. Ivanov }
\affiliation{
 Department of Physics, St. Kliment Ohridski University of Sofia, 5 James Bourchier Boulevard, 1164 Sofia, Bulgaria}

\author{ N. V. Vitanov }
\affiliation{
 Department of Physics, St. Kliment Ohridski University of Sofia, 5 James Bourchier Boulevard, 1164 Sofia, Bulgaria}


\begin{abstract}
In this paper we investigate the quantum dynamics of two spin-1 systems, $\vec{\textbf{S}}_1$ and $\vec{\textbf{S}}_2$, adopting a generalized $(\vec{\textbf{S}}_1+\vec{\textbf{S}}_2)^2$-nonconserving Heisenberg model.
We  show that, due to its symmetry property, the nine-dimensional dynamics of the two qutrits exactly decouples into the direct sum of two sub-dynamics living in two orthogonal four- and five-dimensional subspaces.
Such a reduction is further strengthened by our central result consisting in the fact that in the four-dimensional dynamically invariant subspace, the two qutrits quantum dynamics, with no approximations, is equivalent to that of two non interacting spin 1/2's.
The interpretative advantages stemming from such a remarkable and non-intuitive nesting are systematically exploited and various intriguing features consequently emerging in the dynamics of the two qutrits are deeply scrutinised.
The possibility of exploiting the dynamical reduction brought to light in this paper for exactly treating as well time-dependent versions of our Hamiltonian model is briefly discussed.   
\end{abstract}

\date{\today}


\maketitle

\section{Introduction}
Interacting spin systems with $s>1/2$ reveal a rich variety of phenomena in condensed matter and atomic physics. 
For example, spin models with higher spin length may exhibit novel topological phases described by a hidden order parameter \cite{Haldane1983}. 
Moreover, various strongly interacting spin-boson systems can be mapped onto coupled spin models \cite{Duan2003,Ji2007,Ivanov2014}. 
Apart from the methods used to solve analytically various spin-1/2 systems, in general, models with $s>1/2$ are highly complex and do not permit analytical treatment.

In this paper we consider a system of two interacting spin-1 systems denoted by $\mathbf{S}_1$ and $\mathbf{S}_2$, respectively living in the Hilbert spaces $\mathcal{H}_1$ and $\mathcal{H}_2$, in a physical model described by the time-independent Hamiltonian
\begin{equation} \label{General H}
\hat{H} = \mu_B (g_1 B^z_1 \hat{S}_1^z + g_2 B^z_2 \hat{S}_2^z) + J_0 \hat{\mathbf{S}}_1 \cdot \hat{\mathbf{S}}_2 + \hat{\mathbf{S}}_1 \cdot \mathbf{D}_{12} \cdot \hat{\mathbf{S}}_2.
\end{equation}
The first two terms in Eq.~\eqref{General H} describe the interaction of the two spins with two (generally different) parallel local magnetic fields oriented along the $\hat{z}$-axis, $B_1^z$ and $B_2^z$, with the assumption of scalar $g$-factors, $g_1$ and $g_2$. 
The third term represents the Heisenberg isotropic exchange interaction of coupling strength $J_0$, while the last term, through the second-order traceless Cartesian tensor $\mathbf{D}_{12}$, accounts for symmetric spin-spin anisotropic couplings stemming from the dipole-dipole (d-d) interaction and anisotropic exchange interaction.

A part of the motivation for the present work stems from the growing interest in qutrits -- three-state quantum systems.
Qutrits, and qudits in general, offer numerous advantages over qubits beyond the obvious exponential increase of their Hilbert space.
For example, qutrits allow to construct new types of quantum protocols \cite{Bruckner2002,Molina-Terriza2005} and entanglement \cite{Vaziri2002},
  Bell inequalities resistant to noise \cite{Collins2002},
  larger violations of nonlocality \cite{Kaszlikowski2000},
  more secure quantum communication \cite{Bechmann-Pasquinucci2000,Cerf2002},
  optimization of the Hilbert space dimensionality vs. control complexity \cite{Greentree2004}, and others.
To this end, efficient recipes for manipulation of qutrits \cite{Klimov2003,Vitanov2012} and qudits \cite{Ivanov2006} have been proposed.

Although we only consider two interacting spin-1 systems, our model covers a broad range of physical situations.
For example, in solid state physics the coupling between two molecules, which in their ground state possess a total angular momentum (effective spin) $\mathbf{S}=1$, is described using the Hamiltonian model (\ref{General H}),  with the proviso that spin-orbit effect can be neglected \cite{Bolton}.
An optical lattice of two wells, each containing a single atom of spin 1, provides another possible physical scenario wherein manipulation of the atom-atom coupling constants is within experimental reach \cite{Yip}. 
In addition, the interaction between nanomagnets with a total spin of 1, which is of great interest in quantum computing, is  described by the Hamiltonian model (\ref{General H}) \cite{HeXuLiang}. 
Recently, it was shown that the interaction between two separated nitrogen-vacancy centres in diamond can be described by a Heisenberg spin-1 model \cite{Bermudez2011}.
Moreover, spin-1 models can be realized in a linear ion crystal by using atomic species with three metastable levels driven by laser fields \cite{Cohen2014,Senko2015}.

In this paper we investigate the quantum dynamics generated by the $\mathbf{S}^2$-nonconserving Hamiltonian model (\ref{General H}) where $\mathbf{S}^2 = (\mathbf{S}_1 + \mathbf{S}_2)^2$.
Our main result is that the overall nine-dimensional dynamics may be investigated into two, four- and five-dimensional  subspaces, of well-defined symmetry.
In the 4D-subspace the Hamiltonian $\hat{H}$ may be mapped into the Hamiltonian of two non-interacting spin-${1 \over 2}$ systems. 
The consequences of this remarkable reduction are deeply scrutinised, and in particular, the time evolution of the entanglement between the two qutrits by evaluating the negativity of the compound system.

The paper is organized as follows: In Secs.~\ref{model} and \ref{symmetry} we present the spin-1 model and discuss its symmetry properties. 
In Sec.~\ref{4DS} we study the odd-parity 4D-subspace dynamics of the model and show that it is equivalent to a single spin-$\frac{3}{2}$ system.
Based on this, in Sec.~\ref{negativity} we provide an analysis of the entanglement negativity of the states, which belong to the 4D-subspace. 
In Sec.~\ref{5DS} we discuss the properties of the even-parity 5D-subspace, which is equivalent to a single spin-2 system. 
Finally, in Sec.~\ref{C} conclusive remarks are given together with a possible application of our treatment to Hamiltonian models where the two qutrits are subjected to time-dependent magnetic fields.

\section{The Model}\label{model}

Let's suppose that our system possesses $C_2$-symmetry with respect to the $\hat{z}$ direction. In this case the $\mathbf{D}_{12}$ matrix takes the form
\begin{equation}
\mathbf{D}_{12}=
\begin{pmatrix}
d_{xx} & d_{xy} & 0 \\
d_{yx} & d_{yy} & 0 \\
0 & 0 & d_{zz}
\end{pmatrix}
\end{equation}
and the Hamiltonian (\ref{General H}) may be written as
\begin{eqnarray} \label{Hamiltonian}
\hat{H}&=&
\hbar\omega_{1}\hat{\Sigma}_{1}^{z}+\hbar\omega_{2}\hat{\Sigma}_{2}^{z}
+\gamma_{x}\hat{\Sigma}_{1}^{x}\hat{\Sigma}_{2}^{x}
+\gamma_{y}\hat{\Sigma}_{1}^{y}\hat{\Sigma}_{2}^{y}+\gamma_{z}\hat{\Sigma}_{1}^{z}\hat{\Sigma}_{2}^{z}\notag\\
&&+\gamma_{xy}\hat{\Sigma}_{1}^{x}\hat{\Sigma}_{2}^{y}+\gamma_{yx}\hat{\Sigma}_{1}^{y}\hat{\Sigma}_{2}^{x},
\end{eqnarray}
where the Pauli operators $\hat{\Sigma}_i^k$ ($i=1,2$; $k=x,y,z$) for a spin-1 system are related with the spin-1 operator components as 
\begin{equation}\label{Relations Pauli operators-Angular momentum spin 1}
\hat{S}_i^x = {\hbar \over \sqrt{2}} \hat{\Sigma}_i^x, \quad \hat{S}_i^y = {\hbar \over \sqrt{2}} \hat{\Sigma}_i^y,
 \quad \hat{S}_i^z = \hbar \hat{\Sigma}_i^z.
\end{equation}
The seven real parameters appearing in Eq. (\ref{Hamiltonian}) are given by
\begin{eqnarray}
&&\omega_{1}= \mu_{B} g_{1} B_{1}^{z}, \quad \omega_{2}= \mu_{B} g_{2} B_{2}^{z},\notag \\
&&\gamma_{x} = \dfrac{\hbar^2}{2} (J_0 + d_{xx}), \quad \gamma_{y} = \dfrac{\hbar^2}{2} (J_0 + d_{yy}), \quad \gamma_{z} = \hbar^2 (J_0 + d_{zz}),\notag\\
&&\gamma_{xy} = \dfrac{\hbar^2}{2} d_{xy},\quad \gamma_{yx} = \dfrac{\hbar^2}{2} d_{yx}.
\end{eqnarray}

In this paper we wish to keep the considerations as general as possible, without the restrictions of a specific physical situation.
Hence hereafter we do not attribute any specific symmetry constraints to the real parameters appearing in the Hamiltonian model (\ref{Hamiltonian}).
In this manner, our model includes several models in the literature as special cases.
These include the \emph{XXX} ($\gamma_{x} = \gamma_{y} = \gamma_{z}$), \emph{XXZ} ($\gamma_{x} = \gamma_{y}$) and \emph{XYZ} models for two qutrits subjected to an inhomogeneous magnetic field, generalized with the inclusion of the Dzyaloshinskii-Moriya (DM) interaction ($\gamma_{yx} = - \gamma_{xy}$) \cite{Albayrak}. 
In addition, from our Hamiltonian model one may easily recover a lot of other models, e.g., the \emph{XX} and \emph{XY} models ($\gamma_z = 0$) with (or not) the contribution derived by the DM interaction and (or not) the presence of a homogeneous or inhomogeneous magnetic field, recently taken as starting point for investigating the appearance of thermal entanglement in the system of two interacting qutrits \cite{Joyia,Jafarpour-Ashrafpour,Houa-Lei-Chen}.

\section{Canonical transformation of $\hat{H}$ based on symmetry}\label{symmetry} \label{Section 3}

The following symmetry transformation of $\hat{H}$
\begin{equation}\label{Symmetry Canonical Transformation}
\left\{
\begin{aligned}
\hat{\tilde{\Sigma}}_{1}^{x}=-\hat{\Sigma}_{1}^{x}, \hspace{0.5cm}
\hat{\tilde{\Sigma}}_{1}^{y}=-\hat{\Sigma}_{1}^{y}, \hspace{0.5cm}
\hat{\tilde{\Sigma}}_{1}^{z}=\hat{\Sigma}_{1}^{z}, \\
\hat{\tilde{\Sigma}}_{2}^{x}=-\hat{\Sigma}_{2}^{x}, \hspace{0.5cm}
\hat{\tilde{\Sigma}}_{2}^{y}=-\hat{\Sigma}_{2}^{y}, \hspace{0.5cm}
\hat{\tilde{\Sigma}}_{2}^{z}=\hat{\Sigma}_{2}^{z},
\end{aligned}
\right.
\end{equation}
is canonical, such that $\hat{H}\rightarrow \hat{H}$, which implies the existence of a unitary time-independent operator accomplishing the transformation given by Eq. (\ref{Symmetry Canonical Transformation}) which, by construction, is a constant of motion. 
Because the transformation (\ref{Symmetry Canonical Transformation}) is nothing but a rotation of $\pi$ around the $\hat{z}$-axis of each spin, we can write the unitary operator accomplishing this transformation as 
\begin{eqnarray}\label{Rotation Operator}
\hat{K}&=& e^{i \pi \hat{S}_{1}^{z} / \hbar} \otimes e^{i \pi \hat{S}_{2}^{z} / \hbar} =
e^{i \pi \hat{\Sigma}_{1}^{z}} \otimes e^{i \pi \hat{\Sigma}_{2}^{z}} \notag \\
&& = 1-2\bigl[(\hat{\Sigma}_{1}^{z})^{2}+(\hat{\Sigma}_{2}^{z})^{2}\bigr]+4(\hat{\Sigma}_{1}^{z})^{2}(\hat{\Sigma}_{2}^{z})^{2}.
\end{eqnarray}
The matrix representation of the operator $\hat{K}$ in the ordered standard basis
\begin{equation} \label{Standard Ordered Basis}
\{\ket{11},\ket{10},\ket{1-1},\ket{01},\ket{00},\ket{0-1},\ket{-11},\ket{-10},\ket{-1-1}\}
\end{equation}
is diagonal, 
\begin{equation}\label{MF K}
\hat{K}=
\begin{pmatrix}
1 & 0 & 0 & 0 & 0 & 0 & 0 & 0 & 0\\
0 & -1 & 0 & 0 & 0 & 0 & 0 & 0 & 0\\
0 & 0 & 1 & 0 & 0 & 0 & 0 & 0 & 0\\
0 & 0 & 0 & -1 & 0 & 0 & 0 & 0 & 0\\
0 & 0 & 0 & 0 & 1 & 0 & 0 & 0 & 0\\
0 & 0 & 0 & 0 & 0 & -1 & 0 & 0 & 0\\
0 & 0 & 0 & 0 & 0 & 0 & 1 & 0 & 0\\
0 & 0 & 0 & 0 & 0 & 0 & 0 & -1 & 0\\
0 & 0 & 0 & 0 & 0 & 0 & 0 & 0 & 1\\
\end{pmatrix}\\.
\end{equation}
Equation \eqref{MF K} suggests the possibility of expressing $\hat{K}$ as
\begin{equation}\label{Cos Form of K}
\hat{K} = \cos(\pi\hat{\Sigma}_{\rm tot}^z),
\end{equation}
with $\hat{\Sigma}_{\rm tot}^z = \hat{\Sigma}_{1}^{z} + \hat{\Sigma}_{2}^{z}$ being the total spin of the composed system along the $z$ direction.
Equation (\ref{Cos Form of K}) shows that the constant of motion $\hat{K}$ is indeed a parity operator with respect to the collective Pauli spin variable $\hat{\Sigma}_{\rm tot}^z$, since in correspondence to its integer eigenvalues $M = 2,1,0,-1,-2$, $\hat{K}$ has eigenvalues +1 and -1 depending on the parity of $M$.

The existence of this constant of motion subdivides the 9D Hilbert space of the system into two dynamically invariant and orthogonal subspaces corresponding to the two eigenvalues +1 and -1 of $\hat{K}$.
The subspace relative to $K=1$ ($K=-1$), and then to even (odd) values of $M$, will be hereafter referred to as even- (odd-) parity subspace.
It can be easily seen that the unitary and hermitian operator
\begin{equation}
\hat{\mathbb{U}}=
\begin{pmatrix}
0 & 0 & 0 & 0 & 1 & 0 & 0 & 0 & 0\\
1 & 0 & 0 & 0 & 0 & 0 & 0 & 0 & 0\\
0 & 0 & 0 & 0 & 0 & 1 & 0 & 0 & 0\\
0 & 1 & 0 & 0 & 0 & 0 & 0 & 0 & 0\\
0 & 0 & 0 & 0 & 0 & 0 & 1 & 0 & 0\\
0 & 0 & 1 & 0 & 0 & 0 & 0 & 0 & 0\\
0 & 0 & 0 & 0 & 0 & 0 & 0 & 1 & 0\\
0 & 0 & 0 & 1 & 0 & 0 & 0 & 0 & 0\\
0 & 0 & 0 & 0 & 0 & 0 & 0 & 0 & 1\\
\end{pmatrix}\\,
\end{equation}
transforms the operator $\hat{K}$ as follows
\begin{equation}\label{Ktilde}
\hat{\widetilde{K}} = \hat{\mathbb{U}}^{\dagger} \hat{K} \hat{\mathbb{U}}=
\begin{pmatrix}
-1 & 0 & 0 & 0 & 0 & 0 & 0 & 0 & 0\\
0 & -1 & 0 & 0 & 0 & 0 & 0 & 0 & 0\\
0 & 0 & -1 & 0 & 0 & 0 & 0 & 0 & 0\\
0 & 0 & 0 & -1 & 0 & 0 & 0 & 0 & 0\\
0 & 0 & 0 & 0 & 1 & 0 & 0 & 0 & 0\\
0 & 0 & 0 & 0 & 0 & 1 & 0 & 0 & 0\\
0 & 0 & 0 & 0 & 0 & 0 & 1 & 0 & 0\\
0 & 0 & 0 & 0 & 0 & 0 & 0 & 1 & 0\\
0 & 0 & 0 & 0 & 0 & 0 & 0 & 0 & 1\\
\end{pmatrix}\\.
\end{equation}
As a consequence, by transforming $\hat{H}$ into $\hat{\widetilde{H}}=\hat{\mathbb{U}}^{\dagger} \hat{H} \hat{\mathbb{U}}$, we obtain a new Hamiltonian $\hat{\widetilde{H}}$ 
whose matrix form consists of two blocks, one of dimension 4, related to the eigenvalue -1 of the new constant of motion $\hat{\widetilde{K}}$, and the other of dimension five related to the eigenvalue +1 of $\hat{\widetilde{K}}$, representing the two orthogonal sub-dynamics.
The new Hamiltonian $\hat{\widetilde{H}}$ can be written as
\begin{equation}\label{Htilde projection operators}
\hat{\widetilde{H}}=\hat{P}_{-1}\hat{\widetilde{H}}\hat{P}_{-1}+\hat{P}_{+1}\hat{\widetilde{H}}\hat{P}_{+1},
\end{equation}
where we introduced the hermitian operator $\hat{P}_{-1}$ ($\hat{P}_{+1}$) projecting a generic state of the total Hilbert space $\mathcal{H}=\mathcal{H}_1 \otimes \mathcal{H}_2$ into the $\widetilde{K}$-invariant subspace $\mathcal{H}_-$ ($\mathcal{H}_+$) relative to its eigenvalue -1 (+1) such that $\hat{P}_{-1}\hat{\widetilde{H}}\hat{P}_{-1}$ ($\hat{P}_{+1}\hat{\widetilde{H}}\hat{P}_{+1}$) consists in the upper (lower) block of $\hat{\widetilde{H}}$, or better in a matrix with the same dimension (9) of $\hat{\widetilde{H}}$ but with non vanishing entries only in the upper (lower) four (five) dimensional block.

It is worth noticing that the arguments leading to the possibility of representing the Hamiltonian in accordance with Eq. \eqref{Htilde projection operators} hold their validity even for a more general Hamiltonian model $\hat{H}_{\rm gen}$ obtainable from $\hat{H}$ adding terms commuting with $\hat{K}$, e.g., $(\hat{\Sigma}_1^x)^2$, $\hat{\Sigma}_1^z(\hat{\Sigma}_2^y)^2$ and $\hat{\Sigma}_1^x\hat{\Sigma}_1^y\hat{\Sigma}_2^y\hat{\Sigma}_2^x$,

\be\label{Hgen}
\hat{H}_{\rm gen} = \hat{H} + \text{terms commuting with }\hat{K}.
\ee
However, we confine ourselves to the Hamiltonian model \eqref{Hamiltonian} since it is comparatively more accessible in laboratory and in addition, as we will show in the following sections, it generates interesting quantum dynamical behaviour.

Setting
\begin{align}\label{Positions}
\Omega_{+} &=\omega_{1}+\omega_{2},\notag\\
\Omega_{-} &=\omega_{1}-\omega_{2},\notag\\
\gamma_{1} &=\gamma_{x}-\gamma_{y}-i(\gamma_{xy}+\gamma_{yx}),\notag\\
\gamma_{2} &=\gamma_{x}+\gamma_{y}+i(\gamma_{xy}-\gamma_{yx}),
\end{align}
the $4\times4$ block reads
\begin{equation}\label{MF Htilde 4}
\begin{aligned}
\widetilde{H}_- \equiv
\begin{pmatrix}
\hbar\omega_{1} & \gamma_{2} & \gamma_{1} & 0 \\
\gamma_{2}^{*} & \hbar\omega_{2} & 0 & \gamma_{1} \\
\gamma_{1}^{*} & 0 & -\hbar\omega_{2} & \gamma_{2} \\
0 & \gamma_{1}^{*} & \gamma_{2}^{*} & -\hbar\omega_{1} \\
\end{pmatrix}.
\end{aligned}
\end{equation}
The four states of the original  basis are
\begin{equation} \label{Old spin-1 standard basis states involved in 4x4 block}
\ket{e_1}=\ket{10}, \hspace{0.5 cm} \ket{e_2}=\ket{01}, \hspace{0.5 cm} \ket{e_3}=\ket{0-1}, \hspace{0.5 cm} \ket{e_4}=\ket{-10}.
\end{equation}

The lower block of $\hat{\widetilde{H}}$ is represented by the $5\times5$ matrix
\begin{equation}\label{MF Htilde5}
\begin{aligned}
\widetilde{H}_+ \equiv
\begin{pmatrix}
\hbar\Omega_{+}+\gamma_{z} & 0 & \gamma_{1} & 0 & 0\\
0 & \hbar\Omega_{-}-\gamma_{z} & \gamma_{2} & 0 & 0\\
\gamma_{1}^{*} & \gamma_{2}^{*} & 0 & \gamma_{2} & \gamma_{1}\\
0 & 0 & \gamma_{2}^{*} & -\hbar\Omega_{-}-\gamma_{z} & 0\\
0 & 0 & \gamma_{1}^{*} & 0 & -\hbar\Omega_{+}+\gamma_{z}\\
\end{pmatrix}\\
\end{aligned}
\end{equation}
where the five states of the original basis are
\begin{eqnarray}
&&\ket{e_5} = \ket{11},\hspace{0.5 cm} \ket{e_6} = \ket{1-1},  \hspace{0.5 cm} \ket{e_7} = \ket{00},
\notag\\
&&\ket{e_8} = \ket{-11}, \hspace{0.5 cm} \ket{e_9} = \ket{-1-1}.
\end{eqnarray}
Equation \eqref{Htilde projection operators} implies that the quantum dynamics of two qutrits interacting according to the model of Eq. (\ref{Hamiltonian}) factorizes into an effective spin-${3 \over 2}$ system and an effective spin-2 system.

We note that the mathematical steps leading from Eq. \eqref{Hamiltonian} to Eq. \eqref{Htilde projection operators} reproduce analogous results even if we use, mutatis mutandis, the same Hamiltonian model where qudits systematically substitute the appearing qutrits.
Of course the dimensions of the dynamically invariant subspaces existing in the qudits case strictly depends on the dimension of the qudits Hilbert space.

In the next section we will show that in the case of the qutrits a further aspect of such reducibility of the quantum dynamics of the system emerges, leading to physically transparent and far-reaching consequences.

\section{Four dimensional sub-dynamics}\label{4DS}

The eigenvectors of the Hamiltonian $\hat{\widetilde{H}}_{-}$ (\ref{MF Htilde 4}) may be exactly derived by solving the fourth degree relative secular equation, getting $|\psi_{i}\rangle=\sum_{k=1}^{4}\alpha_{i,k}|e_{k}\rangle$, where the coefficients $\alpha_{i,k}$ are given in Appendix \ref{App 4x4 block eigenvectors} in Eqs A1, A2, A3, A4.
The corresponding eigenvalues are
\begin{equation} \label{4x4 block eigenvalues}
\mathcal{E}_1 = E_{1} + E_{2}, \quad \mathcal{E}_2 = E_{1} - E_{2},\quad \mathcal{E}_3=-\mathcal{E}_{2},\quad \mathcal{E}_4=-\mathcal{E}_{1},
\end{equation}
where
\begin{subequations}\label{E11 and E21}
\begin{align}
&E_{1}=\sqrt{{(\hbar\Omega_{+})^{2} \over 4} + |\gamma_1|^2},\label{E11}\\
&E_{2}=\sqrt{{(\hbar\Omega_{-})^{2} \over 4} + |\gamma_2|^2}.
\end{align}
\end{subequations}
The four eigenvalues of $\hat{\widetilde{H}}_{-}$, in view of Eq. (\ref{4x4 block eigenvalues}), may be obtained summing elements of the two pairs $\{ E_{1}, -E_{1} \}$ and $\{ E_{2}, -E_{2} \}$ in all possible ways.
This circumstance hints that the quantum dynamics of the two qutrits restricted to the four dimensional Hilbert subspace generated by $\ket{e_k}$ with $k = 1,2,3,4,$ is traceable back to that of two effective non-interacting spin-${1 \over 2}$ systems, respectively described by two bi-dimensional traceless Hamiltonians $\hat{H}_1$ and $\hat{H}_2$ with eigenvalues $\pm E_{1}$ and $\pm E_{2}$.

To verify this intuition we search for a mapping between the two qutrits original basis states in \eqref{Old spin-1 standard basis states involved in 4x4 block} and the two spin-${1 \over 2}$ basis, that is $\{ \ket{++}, \ket{+-}, \ket{-+}, \ket{--} \}$, in accordance to which the generic eigenstate $\ket{\psi_k}$ of $\hat{\widetilde{H}}_-$ may be represented as a tensorial product between an eigenstate of $\hat{H}_1$ and an eigenstate of $\hat{H}_2$.
Such a mapping consists simply in
\begin{equation}\label{Spin1-Spin1/2 States Mapping}
\begin{aligned}
\ket{10} & \hspace{0,25cm} \leftrightarrow \hspace{0,25cm} \ket{++},\\
\ket{01} & \hspace{0,25cm} \leftrightarrow \hspace{0,25cm} \ket{+-},\\
\ket{0-1} & \hspace{0,25cm} \leftrightarrow \hspace{0,25cm} \ket{-+},\\
\ket{-10} & \hspace{0,25cm} \leftrightarrow \hspace{0,25cm} \ket{--},
\end{aligned}
\end{equation}
where we define the effective spin-$\frac{1}{2}$ states as $\sigma_{i}^{z}\left|\pm\right\rangle_{i}=\pm\left|\pm\right\rangle_{i}$ with $i=1,2$. Indeed, it's straightforward to show that the sub-dynamics of the two spin-1 systems interacting according to \eqref{Hamiltonian}, related to the $\hat{\widetilde{K}}$-invariant subspace of dimension four characterized by the eigenvalue $\widetilde{K}=-1$, may be reinterpreted as the dynamics of two decoupled effective spin-${1 \over 2}$ systems.
Indeed, we can write $\hat{\widetilde{H}}_-$ as
\begin{equation}\label{4x4 block as two spin-1/2}
\hat{\widetilde{H}}_{-}=\hat{H}_{1} \otimes \hat{\mathbb{1}}_{2}+\hat{\mathbb{1}}_{1} \otimes \hat{H}_{2},
\end{equation}
where we define
\begin{subequations}\label{H1 and H2}
\begin{align}
&&\hat{H}_{1}=\frac{\hbar(\omega_{1}+\omega_{2})}{2}\hat{\sigma}_{1}^{z}+
(\gamma_{x}-\gamma_{y})\hat{\sigma}_{1}^{x}+
(\gamma_{xy}+\gamma_{yx})\hat{\sigma}_{1}^{y},\\
&&\hat{H}_{2}=\frac{\hbar(\omega_{1}-\omega_{2})}{2}\hat{\sigma}_{2}^{z}+
(\gamma_{x}+\gamma_{y})\hat{\sigma}_{2}^{x}-
(\gamma_{xy}-\gamma_{yx})\hat{\sigma}_{2}^{y}.
\end{align}
\end{subequations}
The physical interpretation of this sub-dynamics in terms of two spin-${1 \over 2}$ systems is clear and direct: $\hat{H}_1$ ($\hat{H}_2$) describes a fictitious spin-${1\over 2}$ system immersed in an effective magnetic field $\vec{B}_{1}^{\rm eff}$ ($\vec{B}_{2}^{\rm eff}$) expressible as
\begin{eqnarray} \label{Effective Magnetic Field}
&&\vec{B}_{1}^{\rm eff}= \Bigl( (\gamma_{x}-\gamma_{y}), (\gamma_{xy}+\gamma_{yx}), {\hbar\mu \over 2} (g_1 B_1^z + g_2 B_2^z) \Bigr),\notag\\
&&\vec{B}_{2}^{\rm eff}= \Bigl( (\gamma_{x}+\gamma_{y}), (\gamma_{yx}-\gamma_{xy}), {\hbar\mu \over 2} (g_1 B_1^z - g_2 B_2^z) \Bigr),
\end{eqnarray}
such that we have $\hat{\widetilde{H}}_{-}=\sum_{i=1}^{2}\vec{\sigma}_{i}\cdot \vec{B}_{i}^{\rm eff}$.

Since $\hat{\widetilde{H}}_-$ of Eq.~\eqref{4x4 block as two spin-1/2} describes two decoupled spin-${1 \over 2}$ systems, the eigenvectors of $\hat{\widetilde{H}}_-$ may be written in the following factorized form
\begin{equation} \label{relations to construct the total eigenstates from that of the two single spin-1/2 hamiltonians}
\widetilde{H}_{-} \rightarrow
\left\{
\begin{aligned}
&\ket{\psi_{11}} \otimes \ket{\psi_{21}} \rightarrow \ket{\psi_1}, \\
&\ket{\psi_{11}} \otimes \ket{\psi_{22}} \rightarrow \ket{\psi_2}, \\
&\ket{\psi_{12}} \otimes \ket{\psi_{21}} \rightarrow \ket{\psi_3}, \\
&\ket{\psi_{12}} \otimes \ket{\psi_{22}} \rightarrow \ket{\psi_4},
\end{aligned}
\right.
\end{equation}
where $\{\ket{\psi}_{11}, \ket{\psi}_{12} \}$ ($\{ \ket{\psi}_{21}, \ket{\psi}_{22} \}$) are the eigenvectors of $\hat{H}_1$ ($\hat{H}_2$) given explicitly in Appendix \ref{App 4x4 block eigenvectors}. The corresponding eigenenergies for each state are given by Eq. (\ref{4x4 block eigenvalues}).

We emphasize that the two-qutrit systems may be prepared in a state whose evolution is dominated by one admissible Bohr frequency only exactly mappable in the time evolution of a single spin-${1 \over 2}$ system subjected to an appropriate magnetic field (see Eq. \eqref{Effective Magnetic Field}).
In other words, the quantum dynamics of two qutrits generated by the Hamiltonian (\ref{Hamiltonian}) possesses symmetry properties leading to such a peculiar dynamical behaviour.
We note finally that, since the unitary operator $\hat{\mathbb{U}}$ transforming $\hat{H}$ into the direct sum of $\hat{H}_-$ and $\hat{H}_+$ is independent of time, in view of Ref. \cite{Mess-Nak}, having demonstrated that the quantum dynamics induced by $\hat{H}_-$ may be traced back to that of two effective spin-${1 \over 2}$ systems might provide significant advantages 
even when $\hat{H}$ is time-dependent, at least in its 4D dynamically invariant subspace.

\subsection{States with a specific structure invariant in time}

The odd-parity subspace of both $\hat{H}$ and $\hat{H}_{gen}$ of Eq.~\eqref{Hgen} is spanned by the eigenvectors $\{ \ket{10}, \ket{01}, \ket{0-1}, \ket{-10} \}$ of $\hat{\Sigma}^z_{\rm tot}$. 
Inside such a subspace the two spin-1 systems may be prepared in the following normalized generic superposition of two (generic as well) eigenstates of $\hat{\Sigma}^z_{\rm tot}$ of eigenvalue $M=\pm 1$ peculiarly sharing the same pair of amplitudes in the respective two sub-bases $\{ \ket{10}, \ket{01} \}$ and $\{ \ket{0-1}, \ket{-10} \}$, namely
\begin{equation} \label{Initial state specific structure}
\ket{\Psi(0)} = a \Bigl[ c \ket{10} + d \ket{01} \Bigr] + b \Bigl[ c \ket{0-1} + d \ket{-10} \Bigr],
\end{equation}
where the four complex amplitudes $ac, ad, bc, bd$ fulfil the normalization condition
\begin{equation}\label{Normalization condition special structure}
(|a|^2+|b|^2)(|c|^2+|d|^2) = 1.
\end{equation}
Eqs. \eqref{Initial state specific structure} and \eqref{Normalization condition special structure} individuate a proper subclass of states (initial conditions) sharing a characterizing expansion structure in the standard basis of the odd-parity subspace.

If the quantum dynamics of the two coupled spin-1 systems is governed by a Hamiltonian model $\hat{H}_{gen}$, the initial state given by  Eq. \eqref{Initial state specific structure} would evolve assuming the form
\begin{equation}
\begin{aligned}
\ket{\Psi(t)} = & a(t) \Bigl[ c(t) \ket{10} + d(t) \ket{01} \Bigr] \\
& + b(t) \Bigl[ c'(t) \ket{0-1} + d'(t) \ket{-10} \Bigr],
\end{aligned}
\end{equation}
with $a(0)=a$, $b(0)=b$, $c(0)=c'(0)=c$, $d(0)=d'(0)=d$ and where in general $c(t) \neq c'(t)$ and $d(t) \neq d'(t)$.
Stated another way, it is legitimate to claim that, under $\hat{H}_{gen}$, the two spin-1 systems initial state $\ket{\Psi(0)}$ evolves not preserving its initial structure. This result is not surprising in view of the fact that the four eigenvectors of $\hat{\Sigma}^z_{\rm tot}$ involved in the initial state $\ket{\Psi(0)}$ are not eigenstates of $\hat{H}_{gen}$.
On this basis it appears rather unexpected that the time evolution of $\ket{\Psi(0)}$ determined by the Hamiltonian model \eqref{Hamiltonian} adopted in this paper imposes the special condition $c(t)=c'(t)$ and $d(t)=d'(t)$ at any time instant $t$, namely
\begin{equation} \label{Psi t special structure}
\begin{aligned}
\ket{\Psi(t)} = & a(t) \Bigl[ c(t) \ket{10} + d(t) \ket{01} \Bigr] \\
& + b(t) \Bigl[ c(t) \ket{0-1} + d(t) \ket{-10} \Bigr].
\end{aligned}
\end{equation}
This is indeed an interesting result meaning a sort of time invariance of the structure imposed to the initial state given by Eq. \eqref{Initial state specific structure}.
In the Appendix \ref{App specific structure} we report the explicit expressions of the coefficients $a(t)$, $b(t)$, $c(t)$ and $d(t)$.

The properties possessed by our Hamiltonian model and brought to light in the introductive part of this section provide the basis for an easy interpretation of the result of Eq. \eqref{Psi t special structure}, making transparent how the existence of effective spin-$\frac12$ sub-dynamics is reflected in the quantum dynamics of the two spin-1 systems.
Using the mapping expressed by Eq. (\ref{Spin1-Spin1/2 States Mapping}), the initial state $\ket{\Psi(0)}$ is indeed immediately seen to correspond to the following factorized initial state of the two fictitious spin-1/2 systems
\begin{equation}\label{Special structure state in terms of spin 1/2's}
\ket{\overline{\Psi}(0)} = \Bigl[ a \ket{+}_1 + b \ket{-}_1 \Bigr] \otimes \Bigl[ c \ket{+}_2 + d \ket{-}_2 \Bigr].
\end{equation}
In view of Eq. \eqref{4x4 block as two spin-1/2}, it's then easy to deduce that this  state evolves into the state
\begin{equation}\label{Psi t in terms of spin 1/2}
\ket{\overline{\Psi}(t)} = \Bigl[ a(t) \ket{+}_1 + b(t) \ket{-}_1 \Bigr] \otimes \Bigl[ c(t) \ket{+}_2 + d(t) \ket{-}_2 \Bigr],
\end{equation}
clearly keeping its initial factorization at any time instant $t$.

Looking at Eq. \eqref{Normalization condition special structure} one might wonder whether the state of each effective spin-1/2 systems should be subjected to its own normalization condition, that is whether one should require $|a|^2+|b|^2=1$ together with $|c|^2+|d|^2=1$.
The answer is negative since the only probabilities we are indeed interested in, that is of experimental meaning for our system, are the joint probabilities relative to the two spin-1/2 systems. This consideration of course complies with Eq. \eqref{Normalization condition special structure}.

In order to better appreciate and strengthen the interplay between the quantum dynamics of the two spin-1 systems and that of the two spin-1/2 systems in the odd-parity subspace of the total Hilbert space $\mathcal{H}$, we now evaluate and discuss the time dependence of the mean value of some exemplary and transparent physical observables of the two spin-1 systems on the state $\ket{\Psi(t)}$ given in Eq. \eqref{Psi t special structure}.

It is possible to demonstrate that at any time instant $t$
\begin{subequations}
\begin{align}
& \average{\Psi(t)|\hat{S}_1^z+\hat{S}_2^z|\Psi(t)}=\hbar{|a(t)|^2-|b(t)|^2 \over |a(t)|^2+|b(t)|^2} \label{S_1^z(t)+S_2^z(t) special structure}, \\
& \average{\Psi(t)|\hat{S}_1^z-\hat{S}_2^z|\Psi(t)}=\hbar{|c(t)|^2-|d(t)|^2 \over |c(t)|^2+|d(t)|^2} \label{S_1^z(t)-S_2^z(t) special structure},
\end{align}
\end{subequations}
where $a(t), b(t), c(t), d(t)$ are given in Appendix \ref{App specific structure}.
The right-hand-side expressions of these two equations suggest an interpretation in terms of mean values of appropriate ``physical observables'' related to the two spin-1/2 systems on the state $\ket{\overline{\Psi}(t)}$ given in \eqref{Psi t in terms of spin 1/2}. It is indeed easy to persuade oneself that
\begin{subequations}
\begin{align}
& {|a(t)|^2-|b(t)|^2 \over |a(t)|^2+|b(t)|^2} = \average{\overline{\Psi}(t)|\hat{\sigma}_1^z|\overline{\Psi}(t)},\label{Mean value sigma_1^z special structure state} \\
& {|c(t)|^2-|d(t)|^2 \over |c(t)|^2+|d(t)|^2} = \average{\overline{\Psi}(t)|\hat{\sigma}_2^z|\overline{\Psi}(t)}. \label{Mean value sigma_2^z special structure state}
\end{align}
\end{subequations}
Equation \eqref{Mean value sigma_1^z special structure state} clearly discloses that the temporal behaviour of the total magnetization of the compound spin-1 systems is entirely traceable back to the time dependence of the $z$-component of the first fictitious spin-1/2 systems.
This asymmetry may be fully understood with the help of Eq. \eqref{Psi t special structure} observing that $a(t)$ and $b(t)$ are proportional to the time dependent amplitudes of measuring the maximum and the minimum admissible value of $\hat{S}^z \equiv \hat{S}_1^z+\hat{S}_2^z$ respectively, provided one takes appropriately into account Eq. \eqref{Normalization condition special structure}.
Equation \eqref{Mean value sigma_2^z special structure state} transparently relates the time dependence of $\hat{S}_1^z-\hat{S}_2^z$ to the time evolution of the mean value of the $z$-component of the second fictitious spin-1/2 system.
It is possible to capture the origin of such a result changing the role of $a(t)$ and $b(t)$ with that of $c(t)$ and $d(t)$ respectively which amounts at rewriting Eq. \eqref{Psi t special structure} as follows
\begin{equation}
\begin{aligned}
\ket{\overline{\Psi}(t)} = & c(t) \Bigl[ a(t) \ket{10} + b(t) \ket{0-1} \Bigr] \\
& + d(t) \Bigl[ a(t) \ket{01} + b(t) \ket{-10} \Bigr].
\end{aligned}
\end{equation}
Applying to this equation the same arguments used for interpreting Eq. \eqref{Mean value sigma_1^z special structure state} we easily appreciate the interplay between the mean value of $\hat{S}_1^z-\hat{S}_2^z$ and that of $\hat{\sigma}_2^z$ at any time instant.

It is worth noting that when the amplitudes $a$ and $b$ ($c$ and $d$) appearing in Eq. \eqref{Special structure state in terms of spin 1/2's} are fixed in such a way that $\ket{\overline{\Psi}(0)}$ is an eigenstate of $\hat{H}_1 \otimes \mathbb{1}_2$ ($\mathbb{1}_1 \otimes \hat{H}_2$) then the mean value of $\hat{S}_1^z(t)+\hat{S}_2^z(t)$ ($\hat{S}_1^z(t)-\hat{S}_2^z(t)$) does not evolve in time even if the mean values of $\hat{S}_1^z(t)$ and $\hat{S}_2^z(t)$ evolve in time (unless this special choice is simultaneously made for both the amplitude pairs ($a$, $b$) and ($c$, $d$)).
We in addition emphasize that, once more as a consequence of the sub-dynamics exhibited by $\hat{H}$ in the odd-parity 4D subspace, the mean value of the magnetization of the compound system exhibits sinusoidal oscillations at the frequency ${2 E_{1} \over \hbar}$ where $E_{1}$ is given by Eq. \eqref{E11}.
This behaviour is directly related to Eqs. \eqref{S_1^z(t)+S_2^z(t) special structure} and \eqref{Mean value sigma_1^z special structure state} and occurs whatever the initial state of the two spin-1 systems, as expressed by Eqs. \eqref{Initial state specific structure} and \eqref{Normalization condition special structure}, is (with the exception of the particular initial states previously considered, namely when $\ket{\Psi(0)}$ is mapped into $\ket{\overline{\Psi}(0)}$ eigenstate of $\hat{H}_1 \otimes \mathbb{1}_2$).

In order to illustrate this time dependence of the magnetization let us assume for simplicity that the initial state is given by Eq. \eqref{Initial state specific structure} characterized by equal amplitudes, namely
\begin{equation}\label{Equal weights special structure condition}
a(0)=b(0)=c(0)=d(0)={1 \over \sqrt{2}}.
\end{equation}
Under such condition it is easy to get
\begin{equation}\label{Sz tot special structur equal weights}
\average{\hat{S}^z(t)} = A(\rho_1) \cos \Bigl({1+\rho_1^2 \over \rho_1} \tau-\phi(\rho_1) \Bigr)+C(\rho_1),
\end{equation}
with
\begin{equation}\label{Definitions tau ro}
\begin{aligned}
&\tau={|\gamma_1|t \over \hbar}; \qquad
\rho_1 = {\epsilon_1 \over |\gamma_1|}; \\
&\cos(\phi(\rho_1)) = {-C(\rho_1) \over A(\rho_1)}; \quad \sin(\phi(\rho_1))= {B(\rho_1) \over A(\rho_1)}; \\
&A(\rho_1) = \sqrt{B^2+C^2}; \\
&B(\rho_1) = {2 \rho_1 \over \rho_1^2+1} {\text{Im}[\gamma_1] \over |\gamma_1|}; \\
&C(\rho_1) = {2\rho_1 (\rho_1^2-1) \over \rho_1^2+1} {\text{Re}[\gamma_1] \over |\gamma_1|}.
\end{aligned}
\end{equation}
where $\epsilon_j$ and $\gamma_j$ ($j=1,2$) are defined in Eq. A2 and Eq. \eqref{Positions}, respectively.

{We note that, in view of Eqs. \eqref{Sz tot special structur equal weights} and \eqref{Definitions tau ro} and also taking into account the $\rho$-dependence of $\epsilon_1$, increasing the external parameter $\Omega_+=\omega_1+\omega_2$ which amounts at appropriately acting upon the magnitudes of the two magnetic fields $B_1^z$ and $B_2^z$, the amplitude of the magnetization oscillations increases from 0 to $\sqrt{1-{\text{Re}[\gamma_1] \over |\gamma_1|}}$ when $\rho$ goes from 0 to 1, then decreasing and asymptotically vanishing for large $\rho_1$.
At the same time the frequency of these oscillations goes down to its minimum value 2 (adimensional frequency with respect to the adimensional $\tau$) and then increases asymptotically as $\rho_1$.
It is worth noticing that both the amplitude and the frequency of the magnetization are invariant under the change of $\rho$ with ${1 \over \rho}$.
Looking at $\phi(\rho_1)$ we may instead easily deduce that under the same change of $\rho$ this phase constant undergoes a change of $\pi$.

In Fig. 1 these features characterizing the time behaviour of the magnetization are plotted against the dimensionless time $\tau$ for the following exemplary values of $\rho_1$ while keeping $\gamma_1$ invariant:
\begin{equation}\label{Condition on ro particular case}
\rho_1=10,1,0.1; \qquad \text{Re} [\gamma_1] = {3 \over 5} |\gamma_1| .
\end{equation}

\begin{figure}[tbph]
\centering
{\includegraphics[width=\columnwidth]{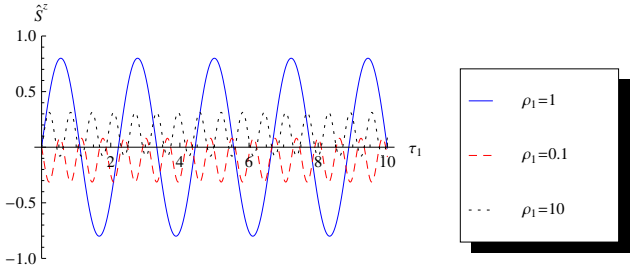} }
\caption{Time dependence of $\average{\hat{S}^z(t)}$ starting from the initial state written in Eq.~\eqref{Initial state specific structure} under the conditions \eqref{Equal weights special structure condition} and \eqref{Condition on ro particular case}.}
\end{figure}

The features exhibited by the time evolution of $\average{\hat{S}^z}$ may be physically understood tracing back to the coincidence existing between such time behaviour and that of the mean value of the Pauli matrix $\hat{\sigma}_1^z$ relative to the first of the two fictitious spin 1/2, as clearly expressed by Eq. \eqref{Mean value sigma_1^z special structure state}.
It is indeed possible to demonstrate that changing the magnetic field $B_1^{\rm eff} \equiv (B_{1x}^{\rm eff},B_{1y}^{\rm eff},B_{1z}^{\rm eff})$ to the related magnetic field
\begin{equation}
\begin{aligned}
\tilde{B}_1^{\rm eff} \equiv \biggl( & \sqrt{{B_{1z}^{\rm eff}-|B_{1}^{\rm eff}| \over 2}},\sqrt{{B_{1z}^{\rm eff}-|B_{1}^{\rm eff}| \over 2}}, \\
&\sqrt{(B_{1x}^{\rm eff})^2+(B_{1y}^{\rm eff})^2}-|B_1^{\rm eff}| \biggl)
\end{aligned}
\end{equation}
realizes in physical terms the change from $\rho$ into ${1 \over \rho_1}$.
At the same time it is possible to show that the first fictitious spin 1/2 driven by $\tilde{B}_1^{\rm eff}$ exhibits a sinusoidal time evolution for the mean value of $\hat{\sigma}_1^z$ coincident with that the first spin 1/2 would have under $B_1^{\rm eff}$, except for the emergence of a difference of phase of $\pi$.

Figure 2 instead reports the time behaviour of $\hat{S}_1^z(t), \hat{S}_2^z(t)$ and $\hat{S}_1^z(t)+\hat{S}_2^z(t) \equiv \hat{S}^z(t)$ assuming the two spin-1 system initially prepared in the following state
\begin{equation}\label{Initial state specific structure with the first spin 1/2 in second eigenstate}
{\epsilon_1 \bigl( \ket{10}+\ket{01} \bigr)+
\gamma_1^* \bigl( \ket{0-1}+\ket{-10} \bigr) \over \sqrt{2 \bigl( \epsilon_{12}^2+|\gamma_1|^2 \bigr)}}
\end{equation}
that is, in terms of the two spin 1/2's
\begin{equation}
\ket{\psi_{11}} \otimes {\ket{+}_2+\ket{-}_2 \over \sqrt{2}}.
\end{equation}
The quantity $\epsilon_1$ and the eigenstate $\ket{\psi_{11}}$ of $\hat{H}_1$ are defined in Appendix \ref{App specific structure}.

Figure 2 displays the time independence of the magnetization in the evolution of the system from its initial state \eqref{Initial state specific structure with the first spin 1/2 in second eigenstate} together with the time dependence of $\hat{S}_1^z(t)$ and $\hat{S}_2^z(t)$ which manifests clearly that the initial state is not a stationary state of $\hat{H}$. The time invariance of $\hat{S}^z(t)$ is certainly traceable back to the stationariness of the first fictitious spin-1/2 system.
\begin{figure}[tbph]
\centering
{\includegraphics[width=\columnwidth]{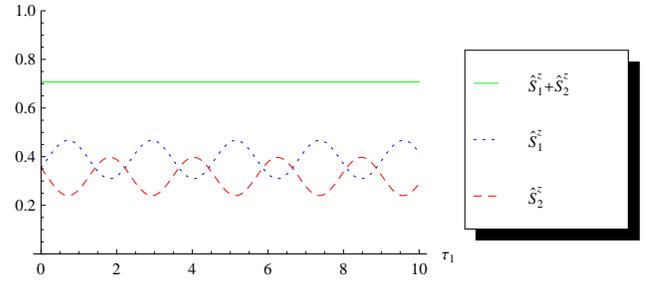} }
\caption{Time dependence of $\average{\hat{S}^z(t)}$ (continue green line), $\average{\hat{S}_1^z(t)}$ (dotted blue line) and $\average{\hat{S}_2^z(t)}$ (red dashed line) starting from the initial state written in Eq.~\eqref{Initial state specific structure with the first spin 1/2 in second eigenstate}.}
\end{figure}
}

\section{Negativity of the two qutrits in their four dimensional subspace}\label{negativity}
In this section we wish to study the negativity, introduced by G. Vidal and R. F. Werner in \cite{Vid-Wer}, possessed by the two qutrits system in a generic pure state belonging to their four dimensional dynamically invariant subspace $\mathcal{H}_-$ and to investigate the emergence of a peculiar behaviour in the time evolution of such a parameter adoptable to measure the entanglement get established between the two spins.
The negativity of a bipartite system constituted by two qudits whose individual Hilbert spaces have dimension $d_1$ and $d_2$ respectively may be defined as \cite{Rai-Luthra}
\begin{equation}\label{General Negativity}
\mathcal{N}_{\hat{\rho}} = {||\hat{\rho}^{T_B} ||_1 - 1 \over d-1},
\end{equation}
where $d=\text{min}\{ d_1,d_2 \}$ and $\hat{\rho}^{T_B}$ is the partial transpose of the matrix $\rho$ representing the state of the total system ($A+B$) with respect to the subsystem $B$. The symbol $|| \cdot ||_1$ is the trace norm which, for a hermitian matrix, is nothing but the sum of the absolute values of its eigenvalues. 
As a consequence the negativity of a state $\rho$ is simply the sum of the absolute values of the negative eigenvalues of ${\rho}^{T_B}$ which is hermitian and such that $\text{Tr}\{ \rho^{T_B} \}=1$.
The value of $\mathcal{N}_{\hat{\rho}}$ ranges from 0 to 1 \cite{Rai-Luthra} and is independent of the factorized orthonormal basis we choose to represent the matrix $\hat{\rho}$. In addition $\mathcal{N}_{\hat{\rho}}$ is independent of the subsystem with respect to which we choose to calculate the partial transpose, given the properties $(\hat{\rho}^{T_A})^T =\hat{\rho}^{T_B}$ and $||X||_1=||X^T||_1$ for any operator $X$.

A generic pure state $\hat{\rho}=\left|\Psi\right\rangle\left\langle\Psi\right|$ belonging to $\mathcal{H}_-$ may be expanded as $\left|\Psi\right\rangle=\sum_{k=1}^{4}c_{k}\left|e_{k}\right\rangle$ ($\sum_{k=1}^{4}|c_{k}|^{2}=1$) in view of Eq. \eqref{Old spin-1 standard basis states involved in 4x4 block}. The corresponding eigenvalues of $\hat{\rho}^{T_{2}}$ are
\begin{eqnarray}
&&\Upsilon_1=1-x,\quad  \Upsilon_2=x,\notag\\
&&\Upsilon_3=\sqrt{x(1-x)},\quad \Upsilon_4=-\Upsilon_3,
\end{eqnarray}
with $x=|c_{1}|^{2}+|c_{4}|^{2}$. Therefore, the negativity of a generic pure state can be written as
\begin{equation}\label{Negativity in H-}
\mathcal{N}_{\hat{\rho}}=\sqrt{x(1-x)},
\end{equation}
which is well defined (since $x \in [0,1]$) and reaches its maximum value $\mathcal{N}_{\hat{\rho}}^{\rm max}=1/2$ at $x=1/2$. 
Thus in the four dimensional dynamically invariant subspace of $\hat{H}$ the negativity exhibited by the two coupled qutrits in a pure state reaches 1/2 as upper limit.
Consequently, the negativity of the two qutrits, since a generic mixed state $\hat{\rho}=\sum_{r=1}^4 p_r \ket{\psi_r}\bra{\psi_r}$ with $\ket{\psi_r}$ in $\mathcal{H}_-$ and ($\sum_{r=1}^4 p_r=1, p_r \geq 0$), possesses the same upper bound 1/2 since \cite{Vid-Wer}
\begin{equation}
\mathcal{N}_{\hat{\rho}} \Bigl(\sum_r \ket{\psi_r}\bra{\psi_r} \Bigr) \leq \sum_r p_r \mathcal{N}_{\hat{\rho}}(\ket{\psi_r}\bra{\psi_r}) \leq {1 \over 2}.
\end{equation}
The existence of a such an upper limit is directly traceable back to the circumstance, easily demonstrable, that every pure state in $\mathcal{H}_-$ possesses a Schmidt decomposition with at most two non-vanishing Schmidt coefficients, namely $k_1$ and $k_2$ expressible as
\begin{equation}
k_1=\sqrt{|c_2|^2+|c_3|^2}, \quad k_2=\sqrt{|c_1|^2+|c_4|^2}, \quad k_3=0.
\end{equation}
When $k_1k_2>0$ the Concurrence ($C(\ket{\psi})$) of two qutrits introduced by Cereceda \cite{Cereceda} reaches its maximum value ${\sqrt{3} \over 2}$ and since in such a case \cite{Rai-Luthra}
\begin{equation}\label{Relation C-N}
{C(\ket{\psi})}=\sqrt{3}\mathcal{N}(\ket{\psi}),
\end{equation}
an upper bound for the Negativity equal to ${1 \over 2}$ emerges in accordance with our previous conclusion.
Thus no pure state in $\mathcal{H}_-$ exhibits maximum entanglement ($C(\ket{\psi})=1$).

It's possible to show that a generic normalized entangled state of the two qutrits in $\mathcal{H}_-$, saturating the negativity at the value $\mathcal{N}_\rho={1 \over 2}$, up to a global phase factor, may be parametrically represented as
\begin{eqnarray}\label{N saturating state}
\ket{\Psi}_{\mathcal{N}}&=& {1 \over \sqrt{2}} \biggl[ \Bigl( \cos(\theta) \ket{1} + e^{i \phi} \sin(\theta) \ket{-1} \Bigr)_1 \otimes \ket{0}_2\notag\\
&& + e^{i \Phi} \ket{0}_1 \otimes \Bigl( \cos(\theta') \ket{1} + e^{i \phi'} \sin(\theta') \ket{-1} \Bigr)_2 \biggr].
\end{eqnarray}
where $\theta$, $\theta'$, $\phi$, $\phi'$ and $\Phi$ freely run in $[0,2\pi]$.

We stress that the existence of the upper bound $1/2$ for the Negativity of a generic pure state belonging to $\mathcal{H}_-$ cannot be extended to $\mathcal{H}$.
It is easy to persuade oneself of this proposition considering the pure normalized state
\begin{equation}\label{Max N state}
{1 \over \sqrt{3}}(\ket{11}+\ket{00}+\ket{-1-1}) \equiv \tilde{k}_1\ket{11}+\tilde{k}_0\ket{00}+\tilde{k}_{-1}\ket{-1-1}
\end{equation}
Since it is in the Schmidt form, we might write its negativity as follows [Rai o Cereceda]
\begin{equation}
\mathcal{N}=\tilde{k}_1\tilde{k}_0+\tilde{k}_1\tilde{k}_{-1}+\tilde{k}_0\tilde{k}_{-1}=1
\end{equation}
which means maximum entanglement.

We point out the possibility that in the parameter space of the Hamiltonian model given by Eq. \eqref{Hamiltonian} there might exist specific examples, built diminishing the number of independent parameters appearing in $\hat{H}$, with eigenvectors in $\mathcal{H}_-$ belonging to the class of states expressed by Eq. \eqref{N saturating state}.
We do not investigate this possibility in its generality confining ourselves to a noticeable example whose four eigenvectors in $\mathcal{H}_-$ are all states of maximum (1/2) negativity.
To this end we claim that, when $\omega_1 = \omega_2$ (homogeneous magnetic field) and $\gamma_{xy} = \gamma_{yx}$ (when the $C_2$-symmetric tensor $\mathbf{D}_{12}$ is symmetric), the four eigenvectors in $\mathcal{H}_-$ of the corresponding Hamiltonian model exhibit $\mathcal{N}_\rho={1 \over 2}$, meaning that each of them may be written in the form \eqref{N saturating state} where $\theta=\theta'$, $\phi$ and $\Phi$ are appropriate expressions of the parameters in $H$.

\subsection{Pure states of $\mathcal{H}_-$ saturating $\mathcal{N}_{\hat{\rho}}$ and evolving with one Bohr frequency only}
In this section we are going to investigate the negativity of a special class of pure states of $\mathcal{H}_-$ by exploiting the advantages stemming from the possibility of describing the four dimensional dynamics of the two spin 1 system in terms of two decoupled spin 1/2's.
We concentrate on the set of pure states of $\mathcal{H}_-$ whose evolution is dominated by one admissible Bohr frequency only. This set consists of any pure state expressible as linear superposition of right two eigenstates of $\tilde{H}_-$ in Eq. \eqref{MF Htilde 4}.
Exploiting the mapping postulated by Eq. \eqref{Spin1-Spin1/2 States Mapping} we are allowed to use the language of the two fictitious spin 1/2's to fully characterize this set of pure states of the two spin 1's.
To this end it is useful to consider two classes of states.
The first one simply encompasses non stationary, normalized and factorized, states of the two spin 1/2's wherein one and only one of them is stationary, that is 
\begin{eqnarray} \label{Factorized state of the two spin-1/2 with one in an eigenstate}
&& \ket{\psi_{11}} \otimes \bigl( \zeta \ket{+}_{2} + \xi \ket{-}_{2} \bigr),\quad \ket{\psi_{12}} \otimes \bigl( \zeta \ket{+}_{2} + \xi \ket{-}_{2} \bigr),\notag\\
&& \bigl( \zeta \ket{+}_{1} + \xi \ket{-}_{1} \bigr) \otimes \ket{\psi_{21}},\quad \bigl( \zeta \ket{+}_{1} + \xi \ket{-}_{1} \bigr) \otimes \ket{\psi_{22}}.
\end{eqnarray}
$\xi$ and $\zeta$ being complex coefficients fulfilling the normalization condition for the two spin 1/2 state.
Besides the states described by Eq. \eqref{Factorized state of the two spin-1/2 with one in an eigenstate} there exist non factorizable, normalized states of the two fictitious spin 1/2's generating as well states of the two spin 1's whose evolution is once more dominated by an admissible Bohr frequency only.
This second class may be represented as follows
\begin{equation}\label{One Bohr non factorized}
\begin{aligned}
& a \ket{\psi_{11}} \ket{\psi_{21}} + b \ket{\psi_{12}} \ket{\psi_{22}}, \\
& a \ket{\psi_{11}} \ket{\psi_{22}} + b \ket{\psi_{12}} \ket{\psi_{21}},
\end{aligned}
\end{equation}
$a$ and $b$ satisfying normalization condition for the compound system.
It is easy to convince oneself that, with the help of Eq. \eqref{Spin1-Spin1/2 States Mapping}, Eqs. \eqref{Factorized state of the two spin-1/2 with one in an eigenstate} and \eqref{One Bohr non factorized} generate all and only the six linear combinations of pairs of stationary states of the two spin 1's, evolving with one Bohr frequency only out of a set of four admissible characteristic Bohr frequencies get in accordance with Eq. \eqref{4x4 block eigenvalues}.
Thus, the language of the two spin 1/2's whose dynamics is governed by the Hamiltonian $\tilde{H}_-$ given by Eq. \eqref{4x4 block as two spin-1/2} provides a simple way for individuating in $\mathcal{H}_-$ all the states of interest in this section with the foreseeable advantage of reducing the quantum dynamics of the two qutrits to that of one or two qubits.

We begin analysing the negativity of the states described by Eq. \eqref{Factorized state of the two spin-1/2 with one in an eigenstate}.
With the help of Eq. \eqref{Spin1-Spin1/2 States Mapping} it's possible to demonstrate that the following condition 
\begin{equation}\label{Condition zeta xi}
|\zeta|^2 = |\xi|^2 = {1 \over 2}
\end{equation}
characterizes the subset of this first class having $\mathcal{N}_\rho=1/2$.
The coefficients $\gamma_1$ and $\gamma_2$ are defined in Eqs. \eqref{Positions}, while $\epsilon_{j}$ may be expressed in terms of the coupling constants and the frequencies appearing in the Hamiltonian model \eqref{Hamiltonian} as follows
\begin{equation}\label{Def epsilon j}
\begin{aligned}
& \epsilon_{j} = \dfrac{\hbar \bigl( \omega_1 - (-1)^j \omega_2 \bigr)}{2} + E_{j}.
\end{aligned}
\end{equation}

The subset fulfilling condition \eqref{Condition zeta xi} may be represented by the following four ($j=1,2$) parametric ($\Phi \in [0,2\pi]$) states of the two spin 1 system
\begin{subequations} \label{General state of class AB}
\begin{align}
\ket{\Psi_{11}} = { \epsilon_{1} \Bigl( \ket{10} + e^{i \Phi} \ket{01} \Bigr)
+ \gamma_1^* \Bigl( \ket{0-1} + e^{i \Phi} \ket{-10} \Bigr)  \over \sqrt{2(\epsilon_{1}^2+|\gamma_1|^2)}}, \label{General state of class AB a} \\
\ket{\Psi_{12}} = { \gamma_2 \Bigl( \ket{10} + e^{i \Phi} \ket{01} \Bigr)
- \epsilon_1 \Bigl( \ket{0-1} + e^{i \Phi} \ket{-10} \Bigr)  \over \sqrt{2(\epsilon_{1}^2+|\gamma_1|^2)}}, \\
\ket{\Psi_{21}} = { \epsilon_{2} \Bigl( \ket{10} + e^{i \Phi} \ket{0-1} \Bigr)
+ \gamma_2^* \Bigl( \ket{01} + e^{i \Phi} \ket{-10} \Bigr) \over \sqrt{2(\epsilon_{2}^2+|\gamma_2|^2)}} \\
\ket{\Psi_{22}} = { \gamma_{2} \Bigl( \ket{10} + e^{i \Phi} \ket{0-1} \Bigr)
- \epsilon_2 \Bigl( \ket{01} + e^{i \Phi} \ket{-10} \Bigr) \over \sqrt{2(\epsilon_{2}^2+|\gamma_2|^2)}}.
\end{align}
\end{subequations}

We emphasize that the existence of a maximum for the negativity of a pure state belonging to $\mathcal{H}_-$ is a property of such a subspace meaning that a pure state which cannot be expanded in the basis given in Eq. \eqref{Old spin-1 standard basis states involved in 4x4 block} might possess negativity higher than 1/2 as it happens for example for the state in Eq. \eqref{Max N state}, orthogonal to $\mathcal{H}_-$.

The time evolution of the negativity of the two spin 1 system prepared in the state $\ket{\Psi_{11}}$ with $\Phi=0$ is reported in Fig. 3 which exhibits a double periodic return to the condition of maximum negativity.
\begin{figure}[tbph]\label{NegMaxInTime}
\centering
{\includegraphics[width=\columnwidth]{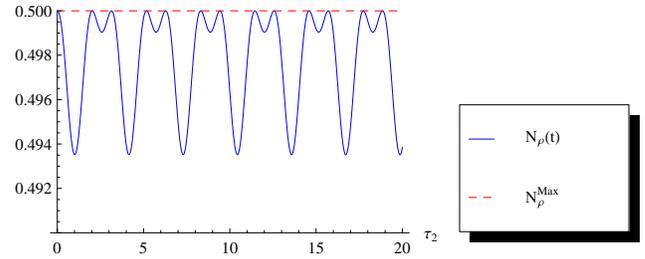} }
\caption{Time evolution of $\mathcal{N}_{\hat{\rho}}(t)$ when the two spin 1 system is initially prepared in the state $\ket{\Psi_{11}}$ in \eqref{General state of class AB a} with $\Phi=0$; the parameter space region is individuated by ${\hbar\Omega_+ \over 2 |\gamma_1|}=1$ and $ \text{Re} [\gamma_1] = {3 \over 5} |\gamma_1|$.}
\end{figure}

To interpret this behaviour, in Fig. 4 we plot $x(t)+1/2$ simultaneously with $|\average{\Psi_{11}(0)|\Psi_{11}(t)}|$ evidencing the periodic return of the system to its initial condition as well as the periodic involvement of another state of maximum negativity.
\begin{figure}[tbph] \label{XandScalarProductMaxNegState}
\centering
{\includegraphics[width=\columnwidth]{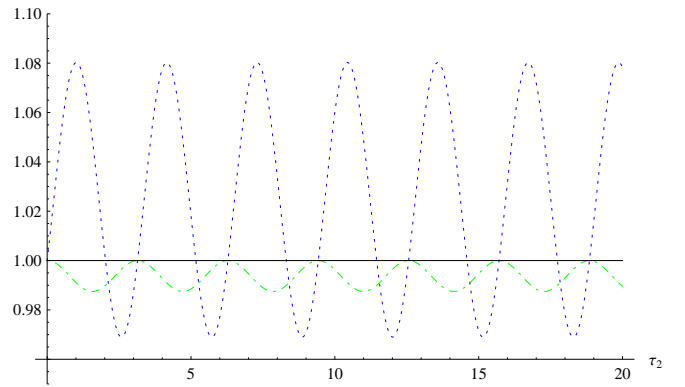} }
\caption{Plot of $x(t)+1/2$ (blue dotted line) and $\average{\Psi_{11}(0)|\Psi_{11}(t)}$ (green dot-dashed line) where $\ket{\Psi_{11}(0)}$ is the initial condition corresponding to $\ket{\Psi_{11}}$ in \eqref{General state of class AB a} with $\Phi=0$ and $\Psi_{11}(t)$ is the related evolved state. The plot shows that at the time instants $\tau_2=k\pi$ ($k=0,1,2,\dots$) the two spin 1 system comes back to its initial condition.}
\end{figure}
It is possible to show that under the condition represented in Fig. 3 the state of $\mathcal{N}_\rho=1/2$, different from the initial one periodically reached by the system, is of the form \eqref{General state of class AB a} with $\Phi \simeq 0.28$.
\begin{figure}[tbph]
\centering
{\includegraphics[width=\columnwidth]{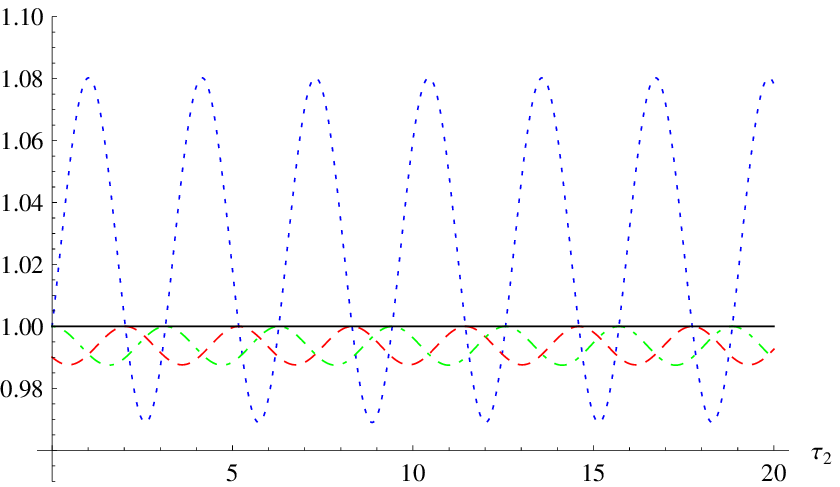} }
\caption{Plot of $x(t)+1/2$ (blue dotted line), $\average{\Psi_{11}(0)|\Psi_{11}(t)}$ (green dot-dashed line) and $\average{\Psi_{11}(\Phi)|\Psi_{11}(t)}$ (red dashed line) where $\ket{\Psi_{11}(0)}$ is the initial condition corresponding to $\ket{\Psi_{11}}$ in \eqref{General state of class AB a} with $\Phi=0$, $\Psi_{11}(t)$ is the related evolved state and $\ket{\Psi_{11}(\Phi)}$ is the state corresponding to $\ket{\Psi_{11}}$ in \eqref{General state of class AB a} with $\Phi \simeq 0.28$. The plot shows that the second point in which $x(t)=\mathcal{N}_{\hat{\rho}}(t)=1/2$ corresponds to the state $\ket{\Psi(\Phi)}$. }
\end{figure}

The states belonging to the first class parametrically expressed by Eq. \eqref{Factorized state of the two spin-1/2 with one in an eigenstate}, besides reaching the maximum admissible value of negativity cannot go below a minimum non vanishing value of the same negativity which can be easily calculated to be
\begin{equation}\label{Minimum Neg}
\mathcal{N}_{\hat{\rho}}^{\rm min}={\epsilon_{j} |\gamma_j| \over \epsilon_{j}^2 + |\gamma_j|^2 } = {\rho_j \over 1 + \rho_j^2}.
\end{equation}
with $\rho_j={\epsilon_j \over |\gamma_j|}$ ($j=1,2$).

This minimum value is assumed in correspondence to the following eight states of the two spin 1/2's
\begin{subequations} \label{Factorizable states of class B with minimum negativity}
\begin{eqnarray}
&& \ket{\psi_{1j}} \otimes \ket{+},\quad \ket{\psi_{1j}} \otimes \ket{-},\label{Factorizable states of class B with minimum negativity a} \\
&& \ket{+}_{1} \otimes \ket{\psi_{2j}}, \quad \ket{-}_{1} \otimes \ket{\psi_{2j}}, \label{Factorizable states of class B with minimum negativity b}
\end{eqnarray}
\end{subequations}
with $j = 1,2$.

The condition on $\zeta$ and $\xi$ generating these states of minimum negativity from the set described by Eq. \eqref{Factorized state of the two spin-1/2 with one in an eigenstate} may be easily expressed in the form $\zeta \xi = 0$ provided $|\zeta|^2+|\xi|^2=1$.

Summing up, the negativity of a generic pure state belonging to the class given by Eq. \eqref{Factorized state of the two spin-1/2 with one in an eigenstate} possesses an upper bound less than 1 (precisely 1/2) and a lower bound, strictly positive, given by Eq. \eqref{Minimum Neg}.

The amplitude of the negativity oscillations exhibited by $\ket{\Psi_{11}}$ is generally $\Phi$-dependent. One might then wonder whether there exist values of $\Phi$ for which such an amplitude reaches its maximum possible value: ${1 \over 2}-{\epsilon_{1} |\gamma_1| \over \epsilon_{1}^2 + |\gamma_1|^2}$, or in other terms, whether the evolved state coincides, up to a global phase factor, with the state $\ket{\psi_{11}} \otimes \ket{\sigma}$ with $\ket{\sigma}=\ket{+}_2$ or $\ket{\sigma}=\ket{-}_2$, in accordance with Eq. \eqref{Factorizable states of class B with minimum negativity}.
We find that the positive or negative answer to such a question depends on the appropriate choice of the region of the parameters space related to the Hamiltonian model depending on the initial state.
 
To appreciate this point we start from the particular couple of states 
\begin{equation}\label{Starting state min neg in time}
\ket{\psi_{11}} \otimes \ket{\pm}_2,
\end{equation}
having minimum negativity and sharing the stationary state of the first fictitious spin 1/2.

We now seek the times at which the negativity restores its initial value as well as the times at which it reaches its maximum value 1/2.
Since we are considering pure states evolving with one Bohr frequency only, the function $x(t)=|c_1(t)|^2+|c_4(t)|^2$ is periodic and then we are sure that the equation $x(t)=x(0)$ admits infinitely many solutions.
This conclusion is valid in every point of the parameter space of the Hamiltonian model.

Concerning whether the system reaches states of maximum negativity we solve the equation $x(t)=1/2$ which assumes, for both states \eqref{Starting state min neg in time}, the same simple form
\begin{equation} \label{Time instant eq.}
\sin^2(\tau_2)={(1+\rho_2^2)^2 \over 8 \rho_2^2} \equiv |s|^2,
\end{equation}
with $\tau_2={E_{2}t \over \hbar}$ and $\rho_2={\epsilon_2 \over |\gamma_2|}$.
We note that in view of Eq. \eqref{Def epsilon j}, the adimensional parameter $\rho_2$ is strictly positive and that Eq. \eqref{Time instant eq.} is solvable only in the $\rho_2$-interval $[\sqrt{2}-1, \sqrt{2}+1]$.
Thus only when the two spin 1's are in a well confined region of the parameter space of the Hamiltonian \eqref{Hamiltonian} the time evolution of the states \eqref{Starting state min neg in time} exhibits negativity oscillations of period $\pi$ from their common $\rho_1$-dependent minimum value to their maximum value (1/2), as displayed in Figs. 6, 7 and 8 where the negativity of both states $\ket{\psi}_{11} \otimes \ket{\pm}$ is reported for exemplary different values of $\rho_2$.
In these plots we fix $\rho_1=1+\sqrt{2}$ determining the same minimum value of the negativity.
\begin{figure}[tbph]
\centering
{\includegraphics[width=\columnwidth]{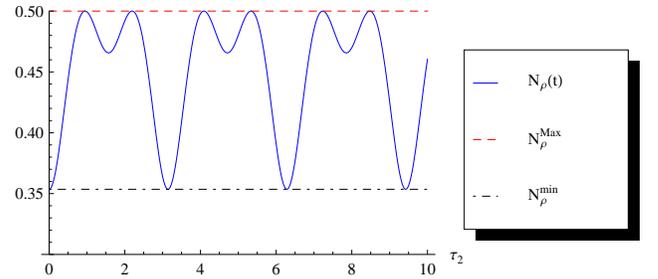} }
\caption{Time evolution of $\mathcal{N}_{\hat{\rho}}$ when the two spin 1 system is initially prepared in the states correspondent to the states $\ket{\psi_{1j}} \otimes \ket{\pm}$ ($j=1,2$) of the two fictitious spin 1/2; the parameters space region is identified by $\rho_1=1+\sqrt{2}$ and $\rho_2 \simeq 1.7$.}
\end{figure}

\begin{figure}[tbph]
\centering
{\includegraphics[width=\columnwidth]{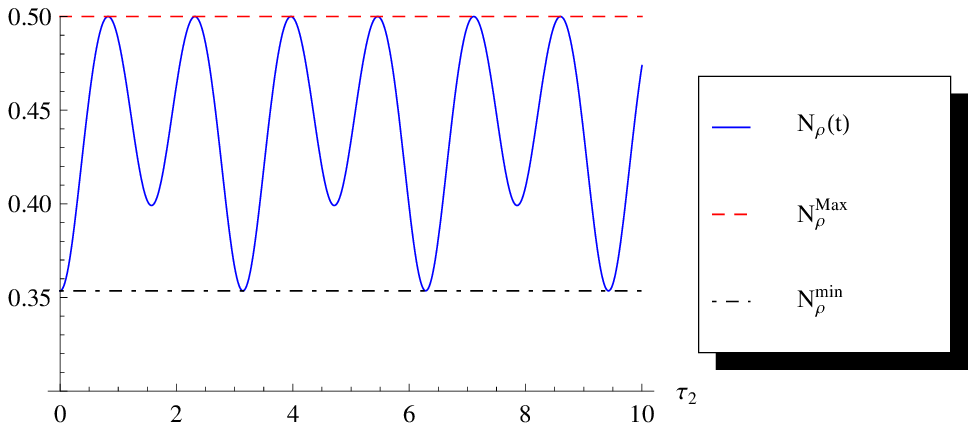} }
\caption{Time evolution of $\mathcal{N}_{\hat{\rho}}$ when the two spin 1 system is initially prepared in the states correspondent to the states $\ket{\psi_{1j}} \otimes \ket{\pm}$ ($j=1,2$) of the two fictitious spin 1/2; the parameters space region is identified by $\rho_1=1+\sqrt{2}$ and $\rho_2 \simeq 1.3$.}
\end{figure}

\begin{figure}[tbph]
\centering
{\includegraphics[width=\columnwidth]{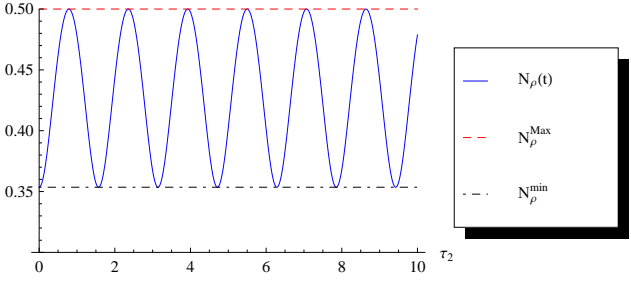} }
\caption{Time evolution of $\mathcal{N}_{\hat{\rho}}$ when the two spin 1 system is initially prepared in the states correspondent to the states $\ket{\psi_{1j}} \otimes \ket{\pm}$ ($j=1,2$) of the two fictitious spin 1/2; the parameters space region is identified by $\rho_1=1+\sqrt{2}$ and $\rho_2 = 1$.}
\end{figure}

The time interval between the two time instants at which $x(t)=1/2$ in any complete oscillation is $\pi-2\arccos(|s|)$ whereas the relative midpoints are $\pi/2 + k\pi$, with $k=0,1,2,\dots$.
It is worth noticing the disappearance of the secondary minima when $\rho_2$ assumes its highest possible value $1+\sqrt{2}$. Moreover, we observe from Figs. 6 and 7 that the closer $\rho_2$ is to 1 and the deeper the secondary minima in the midpoints ($\pi/2 + k\pi$, $k=0,1,2,\dots$) are.
Fig. 8 shows that when $\rho_2=1$ the two spin 1 system recovers its initial negativity with periodicity $\pi/2$.
It is possible to show that with periodicity $\pi$ the two spin 1 system comes back to its initial state too and that, under the special condition $\rho_2=1$, the state of minimum negativity reached at $\pi/2$ is $\ket{\psi}_{11} \otimes \ket{-}$ ($\ket{\psi}_{11} \otimes \ket{+}$) starting from $\ket{\psi}_{11} \otimes \ket{+}$ ($\ket{\psi}_{11} \otimes \ket{-}$).

It is of relevance to observe that the negativity of states \eqref{Starting state min neg in time} \Ignore{for any fixed value of $\rho_1$} is invariant under the change of $\rho_1$ with $1/\rho_1$ and of $\rho_2$ with $1/\rho_2$.
As a consequence the plot 6 coincides with the one referred to $\rho_1 = \sqrt{2}-1$ and $\rho_2 \simeq 1/2$.
This characteristic invariance of the negativity time evolution is a consequence of the way $\hat{H}$ changes under the canonical transformation $\hat{\Sigma}_i^x \rightarrow \hat{\Sigma}_i^x, \hat{\Sigma}_i^y \rightarrow -\hat{\Sigma}_i^y, \hat{\Sigma}_i^z \rightarrow -\hat{\Sigma}_i^z$, describing a rotation of $\pi$ around the $x$-axis.

Had we started from the pair of states of minimum negativity $\ket{\psi}_{12} \otimes \ket{\pm}$ we should have derived exactly the same equation \eqref{Time instant eq.} as well as figures coincident with Figs. 6, 7 and 8.

It's worth noticing that analogous considerations may be developed for the other four states given in Eq. \eqref{Factorizable states of class B with minimum negativity b} with comparable conclusions concerning the time behaviour of the relative negativity.
This time however $\rho_2$ must be replaced by $\rho_1={\epsilon_1 \over |\gamma_1|}$ and the relevant adimensional time becomes $\tau_1={E_{1}t \over \hbar}$.

The time instants at which $\mathcal{N}_\rho=1/2$ satisfies an equation like \eqref{Time instant eq.}, mutatis mutandis, and thus the admissible domain for $\rho_2$ coincides with that of $\rho_1$. 

\Ignore{
analogous conclusions concerning the periodic return to the initial state. Such oscillations exhibit the correspondent maximal excursion of the negativity once again provided to be in a special region of the parameter space of $\hat{H}$.
For example, preparing the two spin 1's in the state $\ket{-}_1 \otimes \ket{\psi_{22}}$, the period of the negativity evolution is $\pi$ with respect to the new adimensional time $\tau_1={E_{1}t \over \hbar}$ as may be determined by solving the equation $x(t)=1/2$ in this case too, that is the equation
\begin{equation} 
\sin^2(\tau_1)={(1+k_1^2)^2 \over 8 k_1^2}
\end{equation}
with $k_1={\epsilon_1 \over |\gamma_1|}$, which should be compared with Eq.\eqref{Time instant eq.}.
Thus, when the adimensional parameter $k_1$ runs in the interval $[\sqrt{3-2\sqrt{2}},\sqrt{3+2\sqrt{2}}]$, as for $k_2$, the negativity evolves up to 1/2 with a period $\pi \over 2$.
}

We stress that, even if the domain of variability of $\rho_1$ and $\rho_2$ is the same, each of them indeed singles out a proper region in the parameter space of $\hat{H}$.
It turns out that these two regions overlap originating Hamiltonian models whose parameters satisfies the domain conditions for both $\rho_1$ and $\rho_2$.
In such a common region of the parameter space it then happens that all the eight states of minimum negativity, as given by Eqs. \eqref{Factorizable states of class B with minimum negativity}, evolve exploring the full domain of the negativity compatible with $\mathcal{H}_-$.

In Figs. 9, 10 and 11 we plot the time evolution of $\mathcal{N}_\rho(t)$ when the system is initially prepared in the states $\ket{\pm} \otimes \ket{\psi_{2j}}$ ($j=1,2$) assuming $\rho_1=1+\sqrt{2}$ and the same three values for $\rho_2$ given it as before.
We notice that, differently from Figs. 6, 7 and 8, Figs. 9, 10 and 11 display a $\rho_2$-dependent level of the minimum negativity and the disappearance of all the secondary minima since $\rho_1=1+\sqrt{2}$ in the three cases.
\begin{figure}[tbph]
\centering
{\includegraphics[width=\columnwidth]{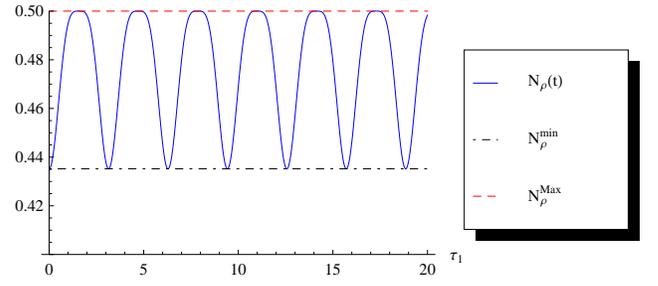} }
\caption{Time evolution of $\mathcal{N}_{\hat{\rho}}$ when the two spin 1 system is initially prepared in the states correspondent to the states $\ket{\pm} \otimes \ket{\psi_{2j}}$ ($j=1,2$) of the two fictitious spin 1/2; the parameters space region is identified by $\rho_1=1+\sqrt{2}$ and $\rho_2 \simeq 1.7$.}
\end{figure}

\begin{figure}[tbph]
\centering
{\includegraphics[width=\columnwidth]{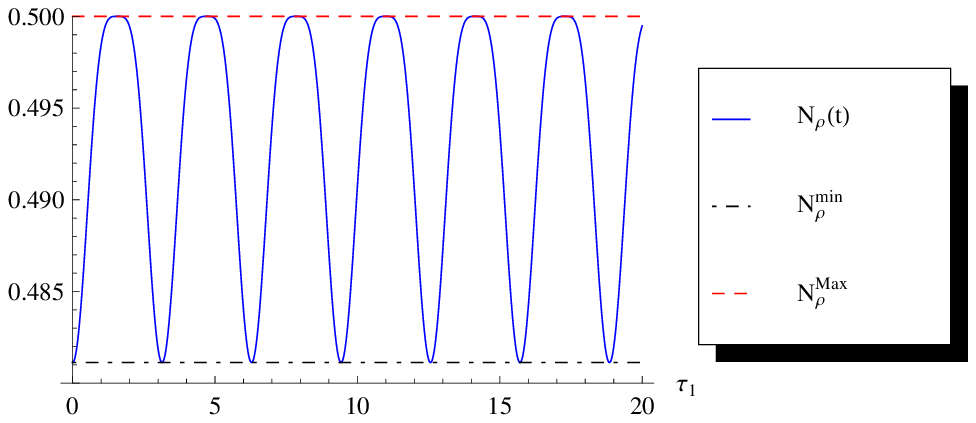} }
\caption{Time evolution of $\mathcal{N}_{\hat{\rho}}$ when the two spin 1 system is initially prepared in the states correspondent to the states $\ket{\pm} \otimes \ket{\psi_{2j}}$ ($j=1,2$) of the two fictitious spin 1/2; the parameters space region is identified by $\rho_1=1+\sqrt{2}$ and $\rho_2 \simeq 1.3$.}
\end{figure}

\begin{figure}[tbph]
\centering
{\includegraphics[width=\columnwidth]{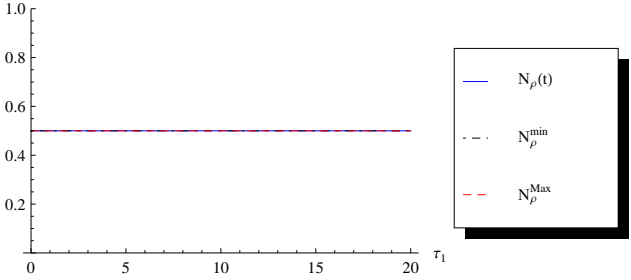} }
\caption{Time evolution of $\mathcal{N}_{\hat{\rho}}$ when the two spin 1 system is initially prepared in the states correspondent to the states $\ket{\pm} \otimes \ket{\psi_{2j}}$ ($j=1,2$) of the two fictitious spin 1/2; the parameters space region is identified by $\rho_1=1+\sqrt{2}$ and $\rho_2 = 1$.}
\end{figure}

Fig. 11 shows the remarkable invariance of the negativity with time when $\rho_2=1$. This fact may be immediately understood observing that, in view of Eq. \eqref{Minimum Neg}, $\mathcal{N}_{\hat{\rho}}^{min}=\mathcal{N}_{\hat{\rho}}^{Max}=1/2$ for all the four states $\ket{\pm} \otimes \ket{\psi_{2j}}$ ($j=1,2$).
We emphasize the time invariance of negativity against the non stationarity of these initial states.

\section{Five dimensional dynamics}\label{5DS}

As shown in Sec.~\eqref{Section 3} the quantum dynamics of the two coupled spin-1 systems is reducible to two quantum sub-dynamics, the first one described by Eq. \eqref{MF Htilde 4} and second one by Eq. \eqref{MF Htilde5}.
The lucky mathematical occurrence leading us to trace back the quantum dynamics of the two spin-1 systems to that of two non interacting spin-1/2 systems in the four dimensional invariant subspace cannot emerge in the other invariant subspace essentially because its dimension 5 is a prime number.
Then, in the spirit of the previous section, the only observation we may do is that in such five dimensional subspace the quantum dynamics of the two spin-1 systems may be mapped into that of a spin-2.
Unfortunately the effective (through an appropriate mapping) representation of $\hat{\widetilde{H}}_5$ in terms of a spin-2 operators is very involved appearing strongly non linear and practically impossible to be related to a convincing physical scenario.
This is why we do not proceed further along this direction confining ourselves to the consideration of particular conditions easily providing the possibility of extracting useful properties possessed by our model.

The first aspect related to the model deserving attention is that by comparing the reduced matrices given by Eqs. \eqref{MF Htilde 4} and \eqref{MF Htilde5} it is possible to note that the parameter $\gamma_{z}$ influences the sub-dynamics in the five dimensional dynamically invariant subspace of $\hat{H}$ only.
As a consequence we may choose specific values of this parameter without modifying the dynamical properties of the system in the four dimensional dynamically invariant subspace.
It is possible to convince oneself that for $\gamma_{z} = 0$ the eigensolutions of $\widetilde{H}_+$ may be exactly found and are given in Appendix \ref{App 5x5 block eigenvectors gammaz=0}.

Furthermore, it is possible to verify that if we assume
\begin{equation}\label{Specific parameters conditions}
\left\{
\begin{aligned}
&\gamma_x = \gamma_y = \gamma \\
&\gamma_{xy} = -\gamma_{yx} = D_z
\end{aligned}
\right.
\end{equation}
the five dimensional block is reduced into two one dimensional block and a three dimensional one as it can be appreciated from what follows
\begin{equation}
\left(
\begin{array}{ccccc}
 \text{$\Omega $}_++\text{$\gamma $}_z & 0 & 0 & 0 & 0 \\
 0 & \text{$\Omega $}_--\text{$\gamma $}_z & 2(\gamma+i D_z)  & 0 & 0 \\
 0 & 2(\gamma-i D_z)  & 0 & 2(\gamma+i D_z) & 0 \\
 0 & 0 & 2(\gamma-i D_z)  & -\text{$\Omega $}_--\text{$\gamma $}_z & 0 \\
 0 & 0 & 0 & 0 & -\text{$\Omega $}_++\text{$\gamma $}_z
\end{array}
\right).
\end{equation}
The previous specific conditions \eqref{Specific parameters conditions} have a clear interesting physical meaning: the first condition imposes an isotropic $XY$-exchange interaction while the second one takes into account the antisymmetric exchange or Dzyaloshinskii-Moriya interaction $\mathbf{D} \cdot (\mathbf{\hat{S}}_1 \times \mathbf{\hat{S}}_2)$ with $\mathbf{D}\equiv(0,0,D_z)$. This model is well known in literature and was studied in connection with the properties of thermal entanglement \cite{Albayrak}.

It is interesting to point out, moreover, that, in this instance, the three dimensional block may be described in terms of the single spin 1 Pauli operators defined in Eq. \eqref{Relations Pauli operators-Angular momentum spin 1} and the relative Hamiltonian precisely reads
\begin{equation}
\hat{\widetilde{H}}_3=
2\gamma \hat{\Sigma}^{x} - 2D_z \hat{\Sigma}^{y}+\Omega_- \hat{\Sigma}^{z} - \gamma_z (\hat{\Sigma}^{z})^{2}.
\end{equation}
We see immediately that putting $\gamma_z=0$ we have a $SU$(2) three dimensional fictitious sub-dynamics of a single spin 1 subjected to the effective external magnetic field $\mathbf{B}_1 \equiv (2\gamma,2D_z,\Omega_-)$, so that we may write $\hat{\widetilde{H}}_3 = \sum_j \mathbf{B}_1^j \cdot \hat{\Sigma}^j$.
This observation is particularly significant at the light of the interplay between the new results obtained for $SU$(2) bidimensional time dependent dynamics \cite{Mess-Nak} \cite{GMN} with the results reported in Ref. \cite{Hioe}.
In this way, under the conditions \eqref{Specific parameters conditions} and $\gamma_{z}=0$, we may study analytically and know exactly the five dimensional sub-dynamics of the two spin 1 systems also in a time dependent scenario, more precisely when the two magnetic fields are time dependent, so that we have $\omega_1(t)$ and $\omega_2(t)$. This target is, however, out of the scope of this paper.

It is worth to point out, finally, that the conditions \eqref{Specific parameters conditions}, contrary to the conditions on $\gamma_{z}$, modify the dynamics in the four dimensional subspace, too.
In this instance, indeed, we obtain a four dimensional sub-dynamics of the two spin 1's well described in terms of two decoupled fictitious spin 1/2's in which the first spin is subjected to a magnetic field only in the $z$-direction while the second spin is immersed in a magnetic field having a direction depending on the coupling parameters of the model.
It can be appreciated and easily verified from Eqs. \eqref{H1 and H2} providing conditions \eqref{Specific parameters conditions}.
Therefore under conditions \eqref{Specific parameters conditions} both the sub-dynamics are exactly treatable or in other words the full model may be exactly solved.

\section{Conclusive remarks and outlooks}\label{C}

In this paper we have examined the quantum dynamics of two spin 1's coupled in accordance with the generalized $\hat{\textbf{S}}^2$-non conserving Heisenberg Hamiltonian model $\hat{H}$ given in Eq. \eqref{Hamiltonian}.
Such a model encompasses a large variety of physical situations and its investigation aims at bringing to light the existence of static and dynamical properties useful to interpret the physical behaviour of a pair of two coupled three state quantum systems, regardless of the specific scenario under scrutiny.

Our first central result is that in each point of the wide parameter space of $\hat{H}$, there exist two dynamically invariant subspaces, having dimensions four and five, which namely are the two only ($\pm1$)-eigenspaces of the parity and constant of motion operator $\hat{K}$ defined in Eq. \eqref{Cos Form of K}.
This subdivision of the total nine-dimensional Hilbert space of the system stems from the invariance of $\hat{H}$ under a rotation of $\pi$ around the $z$-axis.
It unveils that the quantum dynamics of our system breaks into that of two fictitious systems, namely a spin-3/2 and a spin-2 whose quantum evolutions are generated by effective Hamiltonians acting upon own Hilbert spaces isomorphic to the two dynamically invariant subspaces of $\hat{H}$ with parity -1 and +1 respectively.

A careful analysis of the dynamical properties of our system in its four dimensional invariant subspace leads to our second non intuitive and original result.
It consists in the fact that the quantum evolution in such a subspace is exactly traceable back to that of two non interacting fictitious spin 1/2's, subjected to own, fictitious as well, magnetic fields.
Thanks to such a factorization property, one may thus expect to be in condition to interpret and to foresee the time evolution of the two qutrits, prepared in a generic state living in the four dimensional dynamically invariant subspace of $\hat{H}$, exploiting the underlying and simple to evaluate quantum evolutions of two decoupled fictitious spin 1/2's.

This is the reason why most of the paper is dedicated to investigations of properties exhibited by the system in the invariant odd parity subspace, confining in section \eqref{5DS} considerations on the properties of the system evolving in the invariant even parity subspace.

We demonstrate that the interplay between the quantum dynamics of the two spin 1's and that of the compound system of the two fictitious spin 1/2's in the four dimensional subspace $\mathcal{H}_-$ under scrutiny, amounts at the emergence of some dynamical constraints on the time evolution of some classes of states and mean values of observables relative to the two qutrits system.
We show indeed the existence of a class of states of the two qutrits whose given expansion structure in the standard basis of the odd parity invariant subspace keeps such initial structure invariant in time.
Moreover, we find that the time evolution of the total magnetization of the two spin 1's is expressible in a remarkable simple way in terms of the time dependent amplitudes of the evolved state with assigned initial structure.
All these results may be easily interpreted adopting the point of view of the two non interacting spin 1/2's nested in the quantum dynamics of the system restricted in $\mathcal{H}_-$, whose time evolution offers the key leading to a full understanding of the physical origin of the peculiar properties exhibited by our system when it evolves from any state belonging to the class described by Eq. \eqref{Initial state specific structure}.

To investigate whether and how the underlying sub-dynamics of the two non interacting spin 1/2's impose constraints on the time evolution of the entanglement get established between the two qutrits when they are initially prepared in a pure state of $\mathcal{H}_-$ we have studied the negativity bringing to light the existence of an upper limit equal to 1/2.
The origin of such a bound is strictly related to the specific four factorized (standard basis) states of the nine-dimensional total Hilbert space of the two spin 1, generating the four dimensional subspace $\mathcal{H}_-$.
In turn this fact brings to light the role of the $C_2$-symmetry possessed by $\hat{H}$, responsible for the existence of the two specific dynamically invariant subspaces.

It is of relevance to emphasize that our model could be extended even including non linear terms without breaking such $C_2$-symmetry of the Hamiltonian, thus originating once again the subdivision of the total Hilbert space found with our model.
The circumstance that the restriction of $\hat{H}$ into $\mathcal{H}_-$ may be mapped into a model of two non interacting spin 1/2's is instead directly linked to the specific model adopted in this paper, meaning that in presence of appropriate non linear terms, added to our model, the two spin 1/2's would interact.

Once more, we may take advantage of such a mapping also from a dynamical point of view, foreseeing the existence of special classes of non stationary states of the two qutrits manifesting peculiar entanglement properties both static and dynamical.
We are thus led to the four classes of states of $\mathcal{H}_-$, given in Eq. \eqref{Factorized state of the two spin-1/2 with one in an eigenstate}, evolving with an admissible Bohr frequency only showing that each class possesses a lower non vanishing bound of negativity, too, besides the common upper bound.
The paper reports an analysis revealing the region of the parameter space of $\hat{H}$ wherein a state of minimum negativity evolves periodically oscillating between this initial value and 1/2.
In particular we succeed in constructing non stationary states of $\hat{H}$ whose negativity keeps the value 1/2 at any time instant.

In section \eqref{5DS} we have shown that, under specific relations among the the parameters appearing in $\hat{H}$, the underlying dynamics of the effective spin 2 system can be exactly solved.

Recently it has been reported a systematic approach for generating exactly solvable quantum dynamics of a single spin 1/2 subjected to a time dependent magnetic field.
The present work then suggests that the appearance of nested spin 1/2 sub-dynamics in $\mathcal{H}_-$, being treatable in the time dependent case, may led to the construction of exactly solvable time dependent scenarios wherein the two qutrits are subjected to two generally different non-constant magnetic fields.
 
\Ignore{
We have considered the dynamics of two coupled spin-1 systems, which are described by the generalized anisotropic Heisenberg model. 
We have demonstrated that the system allows decomposition into two orthogonal subspaces with well-defined parity. 
The odd-parity subspace corresponds to an effective single spin-$\frac{3}{2}$ system, while the even-parity subspace can be described by an effective single spin-$2$ system. 
We discussed further the decomposition of the $K=-1$ subspace and showed that the hidden sub-dynamics can be captured by introducing two effective uncoupled spin-$\frac{1}{2}$ systems.
{\magenta This decomposition reveals various important features, such as} \black a class of entangled states whose structure remains invariant in time.
Moreover, exploiting the language of the two non-interacting spin-$\frac{1}{2}$ systems we have derived states that have the maximum negativity value and evolve in time under one Bohr frequency only.
Finally, we have shown that for specific values of the initial parameters the underlying dynamics of the effective spin-2 systems can be exactly solved.

Our work suggests that the four-dimensional sub-dynamics, in view of spin-$\frac{1}{2}$ decomposition, can be treated and solved exactly also in the time-dependent case. Furthermore, our model can be extended even including nonlinear terms, which do not break the $C_{2}$-symmetry of the Hamiltonian.
}

\section*{Acknowledgements}
A. M. and R. G. warmly thank Prof. Y. Belousov for commenting the usefulness of the Hamiltonian model and Prof. H. Nakazato for carefully reading the manuscript and for stimulating discussions on the possibility of extending the the treatment to time-dependent scenarios.
Moreover, the same authors acknowledge stimulating conversation on the subject with B. Militello.
R. G. acknowledges kind and stimulating hospitality at the Department of Physics, St. Kliment Ohridski University of Sofia, during the preparation of his master thesis.

\newpage

\appendix
\section{Eigenvectors of the four dimensional dynamics}\label{App 4x4 block eigenvectors}
The eigenvectors of the four dimensional block, in terms of the two spin 1 standard basis states, are
\begin{subequations}\label{4x4 block eigenvectors spin 1}
\begin{align}
& \ket{\psi_{1}}=\dfrac{1}{N} \biggl[ \epsilon_{1} \epsilon_{2} \ket{10}+\epsilon_{1} \gamma_{2}^{*} \ket{01}+\gamma_{1}^{*} \epsilon_{2} \ket{0-1}+\gamma_{1}^{*}\gamma_{2}^{*} \ket{-10} \biggr] \\
& \ket{\psi_{2}}=\dfrac{1}{N} \biggl[ \epsilon_{1} \gamma_2 \ket{10}-\epsilon_{1} \epsilon_2 \ket{01}+\gamma_{1}^{*} \gamma_2 \ket{0-1}-\gamma_{1}^{*}\epsilon_2 \ket{-10} \biggr] \\
& \ket{\psi_{3}}=\dfrac{1}{N} \biggl[ \gamma_1 \epsilon_{2} \ket{10}+\gamma_1 \gamma_{2}^{*} \ket{01}-\epsilon_1 \epsilon_{2} \ket{0-1}-\epsilon_1\gamma_{2}^{*} \ket{-10} \biggr] \\
& \ket{\psi_{4}}=\dfrac{1}{N} \biggl[ \gamma_1 \gamma_2 \ket{10}-\gamma_1 \epsilon_2 \ket{01}-\epsilon_1 \gamma_2 \ket{0-1}+\epsilon_1 \epsilon_2 \ket{-10} \biggr]
\end{align}
\end{subequations}
where the quantity $\epsilon_{j}$ ($j=1,2$) is related to the parameters appearing on the Hamiltonian model \eqref{Hamiltonian} as follows
\begin{equation}
\begin{aligned}
& \epsilon_{j} = \dfrac{\hbar \bigl( \omega_1 - (-1)^j \omega_2 \bigr)}{2} + E_{j}
\end{aligned}
\end{equation}
and $N=N_1 N_2$ is the normalization factor with
\begin{equation}
N_1=\sqrt{\epsilon_1^2+|\gamma_1|^2} \qquad
N_2=\sqrt{\epsilon_2^2+|\gamma_2|^2}
\end{equation}
and finally $\gamma_1$ and $\gamma_2$ are given in Eqs. \eqref{Positions}.

The relative eigenvalues are given in the text in Eqs. \eqref{4x4 block eigenvalues} and \eqref{E11 and E21}.

In view of the reinterpretation of the four dimensional sub-dynamics in terms of two decoupled spin 1/2's, as explicitly made in Eq. \eqref{4x4 block as two spin-1/2}, the four eigenvectors of the dynamics under scrutiny may be written as factorized states of the two fictitious spin 1/2's as given in Eq. \eqref{relations to construct the total eigenstates from that of the two single spin-1/2 hamiltonians} where the eigenstates of single spin 1/2 of $\hat{H}_1$ ($\ket{\psi_{11}}$ and $\ket{\psi_{12}}$) and $\hat{H}_2$ ($\ket{\psi_{21}}$ and $\ket{\psi_{22}}$) are
\begin{subequations}
\begin{align}
& \ket{\psi_{11}} = {\epsilon_{1} \ket{+}_1 + \gamma_1^* \ket{-}_1 \over N_1} \\
& \ket{\psi_{12}} = {\gamma_1 \ket{+}_1 - \epsilon_1 \ket{-}_1 \over N_1} \\
& \ket{\psi_{21}} = {\epsilon_{2} \ket{+}_2 + \gamma_2^* \ket{-}_2 \over N_2} \\
& \ket{\psi_{22}} = {\gamma_2 \ket{+}_2 - \epsilon_2 \ket{-}_2 \over N_2}.
\end{align}
\end{subequations}

\section{Time dependent coefficients of the specific structure non changing in time}\label{App specific structure}

The time dependent coefficients of the state $\ket{\Psi(t)}$ in \eqref{Psi t special structure} take the form
\begin{equation}
\begin{aligned}
& a(t) = a \hspace{0.1cm} k_1^+(t) + b \hspace{0.1cm} k_1^{'+}(t) \qquad b(t) = a \hspace{0.1cm} k_1^-(t) + b \hspace{0.1cm} k_1^{'-}(t) \\
& c(t) = c \hspace{0.1cm} k_2^+(t) + d \hspace{0.1cm} k_2^{'+}(t) \qquad d(t) = c \hspace{0.1cm} k_2^-(t) + d \hspace{0.1cm} k_2^{'-}(t)
\end{aligned}
\end{equation}
where
\begin{subequations}
\begin{align}
& k_j^+(t) = {\epsilon_j^2 e^{-{i \over \hbar} E_{j} t} + |\gamma_j|^2 e^{{i \over \hbar} E_{j} t} \over \epsilon_j^2 + |\gamma_j|^2 } \\
& k_j^-(t) = -2i { \epsilon_j \gamma_j \over \epsilon_j^2 + |\gamma_j|^2 } \sin^2\Bigl( {E_j \over \hbar} t \Bigr) \\
& k_j^{'+}(t) = -2i { \epsilon_j \gamma_j^* \over \epsilon_j^2 + |\gamma_j|^2 } \sin^2\Bigl( {E_j \over \hbar} t \Bigr) \\
& k_j^{'-}(t) = {|\gamma_j|^2 e^{-{i \over \hbar} E_{j} t} + \epsilon_j^2 e^{{i \over \hbar} E_{j} t} \over \epsilon_j^2 + |\gamma_j|^2 }
\end{align}
\end{subequations}
\section{Eigenvectors and Eigenvalues of the five dimensional dynamics}\label{App 5x5 block eigenvectors gammaz=0}
For $\gamma_z = 0$ the secular equation relative to $\widetilde{H}_+$ becomes a bi-quadratic equation and so it can be solved exactly.
In this instance the eigenvectors read ($a/b \leftrightarrow \pm$)
\begin{subequations}
\begin{align}
\ket{\psi_{5}}=&-\gamma_{1}\ket{11}-
\dfrac{\Omega_{+}}{\Omega_{-}}\gamma_{2}\ket{1-1}+
\hbar\Omega_{+}\ket{00}+ \notag \\
&+\dfrac{\Omega_{+}}{\Omega_{-}}\gamma_{2}^{*}\ket{-11}+
\gamma_{1}^{*}\ket{-1-1} \\
\ket{\psi_{6/7}}=&-\dfrac{1}{\gamma_{1}}\Biggl[(\gamma_{x}-\gamma_{y})^{2}+(\gamma_{xy}+\gamma_{yx})^{2}\mp E_{6}(\hbar\Omega_{+}\pm E_{6})+ \notag\\
&\hspace{0cm}+\dfrac{\gamma_{2}\gamma_{2}^{*}\bigl(\mp \hbar\Omega_{+}-E_{6}\bigr)}{\hbar\Omega_{-}-E_{6}}+
\dfrac{\gamma_{2}\gamma_{2}^{*}\bigl(\mp \hbar\Omega_{+}-E_{6}\bigr)}{-\hbar\Omega_{-}-E_{6}}\Biggl]\ket{11}+ \notag\\
&\hspace{0cm}+\dfrac{\gamma_{2}(\mp \hbar\Omega_{+}-E_{6})}{\pm \hbar\Omega_{-}-E_{6}}\ket{1-1}+
\Bigl(\hbar\Omega_{+}\pm E_{2}\Bigr)\ket{00}+ \notag\\
&\hspace{0cm}+\dfrac{\gamma_{2}(\mp \hbar\Omega_{+}-E_{6})}{\mp \hbar\Omega_{-}-E_{6}}\ket{-11}+
\gamma_{1}\ket{-1-1}\\
\ket{\psi_{8/9}}=&-\dfrac{1}{\gamma_{1}}\Biggl[(\gamma_{x}-\gamma_{y})^{2}+(\gamma_{xy}+\gamma_{yx})^{2}\mp E_{8}(\hbar\Omega_{+}\pm E_{8})+ \notag \\
&+\dfrac{\gamma_{2}\gamma_{2}^{*}\bigl(\mp \hbar\Omega_{+}-E_{8}\bigr)}{\hbar\Omega_{-}-E_{8}}+
\dfrac{\gamma_{2}\gamma_{2}^{*}\bigl(\mp \hbar\Omega_{+}-E_{8}\bigr)}{-\hbar\Omega_{-}-E_{8}}\Biggl]\ket{11}+ \notag \\
&+\dfrac{\gamma_{2}(\mp \hbar\Omega_{+}-E_{8})}{\pm \hbar\Omega_{-}-E_{8}}\ket{1-1}+
\Bigl(\hbar\Omega_{+}\pm E_{4}\Bigr)\ket{00}+ \notag \\
&+\dfrac{\gamma_{2}(\mp \hbar\Omega_{+}-E_{8})}{\mp \hbar\Omega_{-}-E_{8}}\ket{-11}+
\gamma_{1}\ket{-1-1}
\end{align}
\end{subequations}
with relative eigenvalues
\begin{subequations}
\begin{align}
&E_{5}=0\\
&E_{6}=-\Biggl\{\hbar^{2}(\omega_{1}^{2}+\omega_{2}^{2})+2\biggl[\gamma_{x}^{2}+\gamma_{y}^{2}+\gamma_{xy}^{2}+\gamma_{yx}^{2}- \notag \\
&\hspace{1.5cm}-\Bigl((\hbar^{2}\omega_{1}\omega_{2})^{2}+4\hbar^{2}\omega_{1}\omega_{2}(-\gamma_{x}\gamma_{y}+\gamma_{xy}\gamma_{yx})+ \notag \\
&\hspace{2cm}+(\gamma_{x}^{2}+\gamma_{y}^{2}+\gamma_{xy}^{2}+\gamma_{yx}^{2})^{2}\Bigr)^{1 \over 2}\biggr]\Biggr\}^{1/2}\\
&E_{7}=-E_{6}\\
&E_{8}=-\Biggl\{\hbar^{2}(\omega_{1}^{2}+\omega_{2}^{2})+2\biggl[\gamma_{x}^{2}+\gamma_{y}^{2}+\gamma_{xy}^{2}+\gamma_{yx}^{2}+ \notag \\
&\hspace{1.5cm}+\Bigl((\hbar^{2}\omega_{1}\omega_{2})^{2}+4\hbar^{2}\omega_{1}\omega_{2}(-\gamma_{x}\gamma_{y}+\gamma_{xy}\gamma_{yx})+ \notag \\
&\hspace{2cm}+(\gamma_{x}^{2}+\gamma_{y}^{2}+\gamma_{xy}^{2}+\gamma_{yx}^{2})^{2}\Bigr)^{1 \over 2}\biggr]\Biggr\}^{1/2}\\
&E_{9}=-E_{8}
\end{align}
\end{subequations}


\end{document}